\documentclass[a4paper,aps,prd,twocolumn,groupedaddress,nofootinbib,floatfix]{revtex4}

\usepackage{graphicx}
\usepackage[USenglish]{babel}
\newlength{\fgwidth}
\newcommand{\G}{\mathcal{G}}
\newcommand{\eps}{\varepsilon}

\begin{document}

\title{Reconstruction of signals with unknown spectra\\in
information field theory with parameter uncertainty}

\author{Torsten A. En{\ss}lin}
\author{Mona Frommert}
\affiliation{Max-Planck-Institut f\"ur Astrophysik, Karl-Schwarzschild-Str. 1, 85741 Garching, Germany}
\date{\today}
\pacs{89.70.-a,11.10.-z,98.80.Es,95.75.-z}

\begin{abstract}
The optimal reconstruction of cosmic metric perturbations and other signals requires knowledge
of their power spectra and other parameters. If these are not known a priori, they have to be measured simultaneously from the same data used for the signal reconstruction. We formulate the general problem of signal inference in the presence of unknown parameters within the  framework of information field theory. 
To solve this, we develop a generic parameter uncertainty renormalized estimation (PURE)  technique. 
As a concrete application, we address the problem of reconstructing Gaussian signals with unknown power-spectrum with five different approaches: (i) separate maximum-a-posteriori power spectrum measurement and subsequent reconstruction, (ii) maximum-a-posteriori reconstruction with marginalized power-spectrum, (iii) maximizing the joint posterior of signal and spectrum, (iv) guessing the spectrum from the variance in the Wiener filter map, and (v) renormalization flow analysis of the field theoretical problem providing the PURE filter.
In all cases, the reconstruction can be described or approximated as Wiener filter operations with assumed signal spectra derived from the data according to the same recipe, but with differing coefficients. 
All of these filters, except the renormalized one, exhibit a perception threshold in case of a Jeffreys prior for the unknown spectrum. 
Data modes with variance below this threshold do not affect the signal reconstruction at all. Filter (iv) seems to be similar to the so called Karhune-Lo\`eve and Feldman-Kaiser-Peacock estimators for galaxy power spectra used in cosmology, which therefore should also exhibit a marginal perception threshold if correctly implemented. We present statistical performance tests and show that the PURE filter is superior to the others, especially if the post-Wiener filter corrections are included or in case an additional scale-independent spectral smoothness prior can be adopted.
\end{abstract}
\maketitle

\section{Introduction}
\subsection{The generic sensing problem}

Reception of a signal is strongly aided by prior knowledge of the signals properties. This is especially true in low signal to noise (S/N) situations, in which proper knowledge can make the difference between recognition of a signal and blindness. Our human senses like vision and hearing are strongly enhanced by our knowledge on the possible signals present in the data-stream entering the human brain. The very same is true for signal reception by artificial sensor systems, since signal knowledge permits us to construct optimal filters, suppressing the noise as far as possible while focusing on the data modes with stronger S/N. If sufficient training data are available, or theoretical reasoning permits us to predict signal properties, optimal filter design is possible and relatively straightforward.

However, there are situations, where such knowledge is not available, or is to be excluded on purpose from the analysis, in order to have a prejudice-free signal reconstruction. In such a situation the required parameters have to be measured simultaneously from the same data which is used for the signal reconstruction. Due to the interdependence of reconstructed signal and parameters, the problem becomes non-trivial and in general non-linear, even if the original inference problem was linear for fixed parameter values.

Let us provide a concrete example in cosmology. The cosmic matter distribution and its imprinted metric fluctuations on large scales can be well approximated to be a Gaussian random field obeying statistical isotropy and homogeneity. Knowledge of the power spectrum of these fields permits us to construct optimal and linear reconstruction filters for data of any linear tracers like the cosmic microwave background, the galaxy distribution (approximatively), or the gravitational lensing signature. For a set of cosmological parameters (e.g. Hubble constant, cosmic matter content, ...) these power spectra are known and can be used. However, the cosmological parameters themselves are not precisely known, and our best knowledge might come from the data-set we are analyzing. Furthermore, if we want to be open for non-standard cosmological scenarios, we might not want to put any prior assumption on the functional form of the power spectrum into our signal reconstruction problem.

Therefore, we need signal reconstruction methods, which are capable of dealing with uncertainties in the parameters of the problem. Such methods would be very useful in many situations, where prior knowledge on signal properties are absent or should be avoided. Some loss in fidelity compared to the case where these parameters are known can be expected, however, such methods can be expected to be flexible and robust due to their generic nature and self-tuning abilities.

For the problem of the reconstruction of the cosmic large-scale structure, the key parameter is the cosmic matter power-spectrum. It is known in the field of signal detection, that a statistical verification of the presence of a signal due to an increase in the data variance is possible well before the signal can be reconstructed itself. 
Thus, a measurement of the signal power-spectrum is already possible while the S/N-ratio is too low for map-making, and is therefore immediately available for filter optimization as soon as the critical S/N-ratio is achieved.

\subsection{Derived filters}\label{sec:derivedfilters} 
The signal reconstruction filters derived in this work can all be regarded as or approximated by an application of a data-dependent Wiener filter operator onto the data, which results in a non-linear transformation of the data. The Wiener filter construction requires the knowledge of the signal covariance, or spectrum, the instrument response and the noise covariance. The signal covariance has to be extracted from the data itself, and therefore introduces a data dependence into the filter. The five filters presented in this work differ in the way the assumed covariance is constructed, due to the different philosophies:
\begin{enumerate}
 \item \textbf{MAP spectrum filter:} The maximum a posteriori (MAP) of the spectrum given the data should be a reasonable guess for the signal spectrum assumed in the Wiener reconstruction.
\item \textbf{Classical map:} The inference problem should be marginalized over all possible power spectra. In doing so, and deriving the classical filter equation by extremizing the resulting effective posterior, a data-dependent Wiener filter is derived, in which an effective spectrum emerges. This spectrum differs in general from the MAP-spectrum.
\item \textbf{Joint MAP filter:} Instead of marginalizing the joint posterior of signal and spectrum and then extremizing it with respect to one of those, we can maximize it with respect to both, leading to the joint MAP filter.
\item \textbf{Critical filter:} 
This filter results if one requires the covariance of the Wiener filter map to exhibit exactly its expected variance, while taking the power loss due to the filter operation into account. 
The critical filter implements accurately the idea behind frequently used power spectrum estimation schemes used in cosmology, like the Karhunen-Lo\`eve (KL, \citep{1947KarhunenK,1978LoeveM, 2002MNRAS.335..887T}) and  Feldman-Kaiser-Peacock (FKP,   \citep{1994ApJ...426...23F}) estimators. 
In case of a Jeffreys prior on the spectral normalisation, it exhibits a marginal perception threshold and
marks the demarcation line between filter with, as the three above, and filter without such a threshold, as the next one.  
\item \textbf{PURE filter:} 
Our ultimate filter would implement the Baysian mean of the signal posterior marginalized over the unknown spectral parameter. 
Only this provides the optimal reconstruction algorithm in the sense of minimizing the
reconstruction error variance. This can only be done by a full field
theoretical treatment which incorporates spectrum-uncertainty effects
correcting for imbalances of the induced errors due to over- and
underestimations of the signal spectrum. 
Here, we incorporate such a correction by virtue of an uncertainty-renormalization calculation. 
The resulting parameter uncertainty renormalized estimation (PURE) filter appears only to be a  Wiener filter in case only an infinitesimal amount of uncertainty is added. The renormalized-optimal spectrum as a fixed point of this uncertainty adding operation is different from the spectra of the other filter. 
In case a finite amount of uncertainty is added, the PURE filter contains corrections terms which can not be described exactly as Wiener filtering.
\end{enumerate}

\subsection{Previous works}

The PURE approach is derived within information field theory (IFT). This deals with the information of data on spatially distributed quantities, and is a statistical field theory. The connection of inference problems and statistical field theories was discovered independently by several authors in cosmology
\cite{1985ApJ...289...10F, 1987ApJ...323L.103B, 2009PhRvD..80j5005E}, statistical field theory
\cite{1987PhRvL..58..741B, 1988PhRvL..61.1512B, 1996PhRvL..77.4693B}, and quantum mechanics
\cite{2000FBS....29...25L, 2000PhRvL..84.2068L,
2000PhRvL..84.4517L, 2000PhLA..276...19L, 
2001EPJB...20..349L, 2005EPJB...46...41L}. A pedagogical introduction into IFT can be found in \cite{2009PhRvD..80j5005E}.

The uncomfortable dependence of information theoretical methods on signal prior information have lead several authors to think about methods to extract this information at least partly from the data.  
For example a smoothness prior for the signal can be used, where an ``optimal'' value for the smoothness controlling parameter derives from the data themself
\cite{1996PhRvL..77.4693B}. 
The optimal smoothness constraint for a Gaussian signal is provided by its covariance, as known from Wiener filter theory \citep{1949wiener}. A natural proposal is therefore to measure the power spectrum (or any characteristics of the signal covariance) from the data and to use this for Wiener filtering or other signal reconstruction methods \citep{1987AJ.....93..968R, 1992ApJ...398..169R, 1996ITSP...44.1469L, 1999ISPL....6..205S}. Data gaps complicate the power spectrum measurement step, but  extensions of such methods to this case exist \citep{2000AJ....120.2163S}. However, a more theoretical understanding of the inference problem and the assumptions implicitly made by these methods would be beneficial to answer several questions.
How should the spectrum be measured  optimally? How can spectral prior information be incorporated into the filter? And is the best spectral estimator really the best choice for the spectrum assumed in the Wiener filter?

Only Bayesian approaches, which are explicitely dealing with all relevant prior information, can answer these questions accurately. For example, it is possible to use the MAP approach to the problem of Wiener filtering if the overall amplitude of the signal covariance is unknown, even on a logarithmic scale \cite{1986WRR....22..499K}. For a white signal, where all pixels are statistically independent, this can be generalized to the case that all pixels amplitudes are drawn from a scale-free distribution function \cite{2008ApJ...675.1304R}. 

In precision cosmology, the problem of inferring the image and its power spectrum simultaneously is very prominent in cosmic microwave studies and cosmography of the large scale structure. 
It has been addressed rigorously via the Gibbs sampling scheme \citep{2004PhRvD..70h3511W, 2004ApJS..155..227E, 2004ApJ...609....1J, 2010MNRAS.406...60J}. 
Since this approach samples the full joint posterior of maps and spectra, it provides the full solution to the problem. 
However, the computational costs of Gibbs sampling are high. Also obtaining analytical insights into the general behavior of the scheme is not trivial. 
Computationally cheaper and analytically simpler, or even just alternative methods are therefore interesting and and some of the algorithms provided by this work are good candidates for being this.

\subsection{Structure of the work}

We introduce IFT with parameter uncertainties in Sec.\ \ref{sec:IFTPU}. 
In Sec.\ \ref{sec:ssu} the problem of signal spectrum uncertainty is introduced, and the four of the mentioned filters are derived from MAP principles.
To go beyond the MAP approximation the generic PURE approach is developed in Sec.\ \ref{sec:uncertainty renormalization flow}, where for any case with fourth order interaction terms the generic uncertainty renormalization flow equation is provided. 
The specific application of this approach is given in Sec. \ref{sec:applyingPURE}, where the PURE filter for the problem of reconstruction without spectral knowledge is derived.
The perception thresholds of all these filters are investigated in Sec.\ \ref{sec:perception threshold}, and their fidelity in Sec.\ \ref{sec:compare}, where also a PURE filter with spectral smoothness prior is presented. Finally, we conclude in Sec.\ \ref{sec:conclusion}.

\section{Information field theory with parameter uncertainty}\label{sec:IFTPU}
\subsection{Information field theory}
\label{sec:ift}

We briefly introduce the concepts of IFT and extend them to the case of parameter uncertainties.
A more pedagogical introductions, as well as more details on terminology and notation of the framework can be found in \cite{2009PhRvD..80j5005E}. 
An information field is simply a spatially extended signal, where a signal $s$ is any quantity a scientist might be interested in measuring.  We treat the signal $s(x) = s_x$, a function of a spatial coordinate $x$,  as an abstract vector in Hilbert space with the scalar product
 $j^\dagger s = \int  dx \, \overline{j(x)}\, s(x).$

The goal of IFT is to make statements on the signal field, which is constrained by prior knowledge and observational data. 
Since we are usually dealing with a finite number of noisy data points, a precise reconstruction of a signal field with its infinite number of degrees of freedom is rarely possible. 
Our aim is therefore to investigate the probability function of $s$ given the data $d$, the so called posterior $P(s|d)$.
The posterior is usually constructed from the signal prior $P(s)$ and the likelihood of the data $P(d|s)$ using Bayes theorem
\begin{equation}
 P(s|d) = \frac{P(d|s)\, P(s)}{P(d)}. 
\end{equation}
The normalisation constant here, the so called evidence $P(d)$, is given by a marginalization of the signal field
\begin{equation}
 P(d) = \int \! \mathcal{D}s\, P(d,s),
\end{equation}
where $P(d,s)=P(d|s)\, P(s)$ is the joint probablity density function of data and signal.
The phase space or path integral $\int \! \mathcal{D}s$ goes over all possible signal field configurations, weighted with  $P(d, s)$. 

In IFT, we rewrite Bayes theorem in the language of a statistical field theory, namely as
\begin{equation}
 P(s|d) = \frac{e^{-H[s]}}{Z}, 
\end{equation}
where the information Hamiltonian $H[s] = - \log P(d,s)$ and the partition function  $Z = P(d)$ are actually only a renaming of (the negative logarithm of) the joint probability and evidence. 
This change in language, however, permits to transfer many results from statistical field theory to tackle IFT problems.

The goal of an IFT analysis could be to calculate moments of the signal field averaged in a similar path integral over the posterior $P(s|d)$, e.g. in order to know the mean signal
\begin{equation}
 m = \langle s \rangle_{(s|d)} = \int \! \mathcal{D}s \, s \, P(s|d) .
\end{equation}
This mean is of special interest, since it is optimal in an $\mathcal{L}^2$-error norm sense. 
It minimizes the expected error variance $\langle (s-m)^\dagger (s-m)\rangle_{(s|d)} $ among all possible $m$.

In practical applications, we often discretize the signal field in $N_\mathrm{pix}$ pixels at locations $x_i$. 
Then the discretized path integral for any signal function $f(s)$ is
\begin{displaymath}
 \int \! \mathcal{D}s\, f(s) = \left( \prod_{i=1}^{N_\mathrm{pix}} \,  \int \! ds(x_i) \right) f(s({x_1}), \ldots, s({x_{N_\mathrm{pix}}})).
\end{displaymath}
If possible, we try to avoid to evaluate such very high dimensional integrals nummerically. 
We use the fact that a multimodal Gaussian probability density function as given by 
\begin{equation}\label{eq:GaussPrior}
\G(s,S) \equiv \frac{1}{|2\pi\, S|^\frac{1}{2}}\, \exp\left( -\frac{1}{2} s^\dagger S^{-1} s \right) 
\end{equation}
(with $|S|$ denoting the determined of the matrix $S$) can be integrated analytically: $ \int \! \mathcal{D}s\,\mathcal{G}(s,S) = 1$.
Many functional integrals can be derived from this, like the moments of a Gaussian, and path-integrals of any quadratic functional of the integrated field. 
Non-quadratic exponents can be expanded around the multivariate Gauss integral in terms of diagramatic pertubation series. 
For further details, the reader is refered to \cite{2009PhRvD..80j5005E} and any standard book on field theory.

In the simplest case of the theory, signal and noise are independent Gaussian random variables, and the data depend linearly on them. This so-called free theory can be treated analytically and is our starting point. 
It has been analyzed in depth before and leads to the so called Wiener-filter theory \citep{1949wiener}. 
However, usually the assumption that all parameters $p$ of the problem like instrument calibration, or signal covariance, are known is used. 
This assumption will be dropped in the following, and we will see, that the otherwise trivial case gets interesting complications and the corresponding free IFT is enriched by interaction terms.

\subsection{Free theory from a Gaussian data model}

We assume that the signal we want to reconstruct is a Gaussian random field, with 
a probability distribution prior to any measurement described as $P(s|p) = \G(s,S_p)$,
where $S_p=\langle s\,s^\dagger \rangle_{(s|p)}$ is the signal covariance given the parameter $p$, which itself might be a vector or even a field over some space. The subscript $(s|p)$ on the brackets of the expectation value indicate that the average should be done over the probability distribution $P(s|p)$.
Thus, the individual elements of the signal covariance matrix read
\begin{displaymath}
 (S_p)_{xy} =\langle s(x)\,\overline{s(y)} \rangle_{(s|p)} = \int \! \mathcal{D}s\, s(x) \overline{s(y)}\, P(s|p).
\end{displaymath}

We further assume that the signal is processed by a linear measurement device with response matrix $R$ and additive noise $n$ according to:
\begin{equation}
d = R\, s + n.
\end{equation}
In general, response and noise can also depend on unknown parameters and the general theory developed here can also be applied to that case. To focus the discussion, we only consider here the concrete example of a parameter dependent signal covariance, and assume the response and noise statistics to be known.
We assume the noise to be signal-independent and Gaussian, and thus
\begin{equation}
P(n|s,p) = \G(n,N),
\end{equation}
where $N= \langle n\, n^{\dagger} \rangle_{(n)}$ is the noise covariance matrix. Since the noise is just the difference of the data to the signal-response, $n=d-R\,s$, the likelihood of the data is
\begin{equation}
\label{eq:PsPosterior}
P(d|s,p) = P(n= d-R\,s|s,p) = \G(d-R\,s,N).
\end{equation}

The information Hamiltonian as defined in \cite{2009PhRvD..80j5005E} is the negative logarithm of the joint probability function of signal and data for given and fixed parameters:
\begin{equation}
 H_{p}[s] = - \log P(d,s|p) = -  \log\left[ P(d|s,p) \,P(s|p)\right].
\end{equation}
Thus the Hamiltonian of the Gaussian theory,
\begin{eqnarray}\label{eq:freeHamiltonian}
H_{p}^\G[s]  =
\frac{1}{2} s^{\dagger} D_p^{-1} s - j^{\dagger} s + H_{0,p}^{\G},
\end{eqnarray}
is only quadratic in the signal, and therefore corresponds to a free field theory. Here
\begin{equation}\label{eq:D}
D_p= \left[ S^{-1}_p + M \right]^{-1},\; \mbox{with}\; M= R^{\dagger} N^{-1} R,
\end{equation}
is the information propagator, which depends on the unknown spectral parameters. The information source,
\begin{equation}
j = R^{\dagger} N^{-1} d,
\end{equation}
depends linearly on the data in a response-over-noise weighted fashion. Finally, 
\begin{equation}
 H_{0,p}^{\G} = \frac{1}{2}\, d^\dagger\,N^{-1}\,d + \frac{1}{2}\,\log\left( |2\,\pi\,S_p|\,|2\,\pi\,N| \right) 
\end{equation}
absorbs all $s$-independent normalization constants. It can not be ignored here, since it depends on $p$. 

The key quantity, from which all relevant moments of the signal can be estimated, is the partition function,
\begin{equation}
Z_p[J] =
\int \!\mathcal{D}s \, e^{-H_p[s] + J^{\dagger}s}.
\end{equation}

For the free field theory the partition function is 
\begin{equation}
\label{eq:ZdfreeTheory}
Z_p^{\G}[J] = \sqrt{|2\pi\, D_p|}\,
\exp\left\lbrace +\frac{1}{2} (J+j)^{\dagger} D_p (J+j) -H_{0,p}^{\G} \right\rbrace\!.
\end{equation}
This explicit formula permits us to calculate the expectation of the signal given the data (and the parameters), in the following called the map $m_{p}$:
\begin{eqnarray}
\label{eq:WFmap} 
m_{p} &=& \langle s \rangle_{(s|d,p)}  = \left. \frac{\delta \log Z_p^{\G}}{\delta J}\right|_{J=0} 
=D_p\,j \nonumber\\
&=& \underbrace{\left[ S_p^{-1} + R^{\dagger} N^{-1} R\right]^{-1} R^{\dagger} N^{-1}}_{F_{p}} d.
\end{eqnarray}
The last expression shows that the map is given by the data after applying a generalized Wiener filter, $m_{p} = F_{p} \,d$, which depends on the parameter $p$ of the signal covariance. 

Similarly, the quadratic uncertainty of the signal map can be worked out. It turns out that for a free theory it is the propagator itself
\begin{equation}\label{eq:mapuncertainty}
\langle (s-m_p) (s-m_p)^{\dagger}\rangle_{(s|d,p)} = 
\langle s \, s^{\dagger}\rangle_{(s|d,p)} - m_p\,m_p^\dagger =
D_p .
\end{equation}
The first identity follows from $\langle s\,m_p^\dagger\rangle_{(s|d,p)} = \langle s\rangle_{(s|d,p)} \,m_p^\dagger = m_p m_p^\dagger$ due to the fact, that the reconstructed map $m_p$ is solely determined by the data, and therefore given in this average. The second identity holds due to the identity of the connected correlation function and the propagator, $\langle s\, s^\dagger \rangle^\mathrm{c}_{(s|d,p)} = \delta^2 \log Z_p^\G /\delta Z^2|_{J=0} = D_p$.

\subsection{Classical field theory}

In case of the free theory, the map, Eq.\ \ref{eq:WFmap},  would also be obtained from 
a classical treatment of the Hamiltonian by extremizing it:
\begin{equation}
 \frac{\delta H_p[s]}{\delta s} = 0.
\end{equation}
For a Hamiltonian with interaction terms the classical field (in field-theoretical language) or MAP estimator (in signal processing language) is a useful approximation to the correct expectation value. The inverse Hessian in the signal Hilbert space around this map, 
\begin{equation}
\left( \frac{\delta^2 H_p[s]}{\delta s \, \delta s^\dagger}\right)^{-1},
\end{equation}
characterizes the uncertainty. For the free theory, this is the propagator, as given by Eq.\ \ref{eq:mapuncertainty}.

The identity of fully field theoretical and classical results holds only for the case of a free theory. However, the latter is often an acceptable approximation to the former, while much easier to derive. Therefore, we will make also use of the classical approximation in the following.

\subsection{Parameter uncertainty and posterior}

In many applications, there are parameters specifying the likelihood and prior, and thereby the coefficients of the
Hamiltonian, which are not precisely known. These parameters, in the following denoted by the abstract vector $p$, are either to be determined from the data, to be marginalized over, or to be simultaneously determined with the signal. 

In such a case we have to construct the joint posterior of the signal and the parameter given the data. This is given according to Bayes' theorem as
\begin{eqnarray}
 P(s,p|d) &=&\frac{P(d,s,p)}{P(d)} = P(s|d,p) \, \frac{P(d|p)}{P(d)}\, P(p)
\,,
\end{eqnarray}
where we had to introduce the parameter prior $P(p)$. The last expression contains a Bayes factor $P(d|p)/P(d)$, the ratio of the evidence of data for a specific parameter set to that of the model at all. Thus, the joint posterior is weighted towards model-parameters for which the data provide larger evidence in addition to any prior-weighting.

The definition of the Hamiltonian for fixed-parameters as $H_p[s] =  - \log P(d,s|p)$ 
permits us to construct the joint partition function
\begin{eqnarray}
 Z[J,K] \!\!&\equiv &\!\!  \!\int\! dp \!\int \!\mathcal{D}s \, P(s,p|d)\,P(d)\, e^{J^\dagger s + K^\dagger  p}\\
\!\!&=&\!\! \!\int\! dp\,P(p)\, e^{K^\dagger p} \!\underbrace{\int \!\mathcal{D}s \,\overbrace{P(d|s,p)\,P(s|p)}^{P(d,s|p)=e^{-H_p[s]}}\,e^{J^\dagger s}}_{Z_p[J]},\nonumber
\end{eqnarray}
which is built upon $Z_p[J]$, the partition function of the theory for given parameters $p$. The information field estimators, marginalized for the unknown parameters, is then simply given by
\begin{eqnarray}\label{eq:m_eff}
 \langle s \rangle_{(s|d)} &=& \left. \frac{\delta\,\log Z[J,K] }{\delta J} \right|_{J,K=0} \nonumber\\
&=& \left. \frac{1}{Z} \, \int dp\,P(p) \, \frac{\delta Z_p[J]}{\delta J}\right|_{J=0}\\
&=& \int dp\,\underbrace{P(p)\, \frac{Z_p}{Z}}_{P(p|d)}\,   \langle s \rangle_{(s|d,p)}.  \nonumber
\end{eqnarray}
The aim of this work is to provide schemes to calculate this parameter marginalized signal mean.
It is not just the signal estimator multiplied by the parameter prior $P(p)$, but is additionally weighted by a parameter likelihood factor $P(d|p) =  Z_p$, so that the parameter-dependent signal means are averaged over the parameter posterior $P(p|d)$. 
Therefore, parameter values which are especially compatible with the data get automatically a larger weight, as recognized before.

\subsection{Effective marginalized Hamiltonian} \label{sec:effectiveH}

If a parameter-dependent Hamiltonian $H_p[s] = - \log P(d,s|p)$ describes the conditional probability of the signal and data given the parameters, an effective, parameter-marginalized Hamiltonian $H[s]$ is defined by
\begin{eqnarray}\label{eq:effectiveHamiltonianGeneral}
 e^{-H[s]} &\equiv& \int \!\! dp\,  P(d,s,p)  
= \int \!\! dp\, P(d,s|p)\, P(p) \nonumber\\
&=& 
 \int \!\! dp \, e^{-H_p[s]- E_p}  \,,
\end{eqnarray} 
with $E_p = -\log P(p)$ the parameter-prior-energy.
It is crucial, that $H_p[s]$ obeys the correct normalization condition, $\int \!\! \mathcal{D}d \,\int \!\! \mathcal{D}s \, \exp(-H_p[s])=1$, otherwise a hidden prior on $p$ may enter the calculation.

In many cases, an analytical calculation of the effective Hamiltonian will be
out of reach. Since the perturbative field theoretical treatment requires a
polynomial representation anyway, it is often easier to obtain the coefficients
of the effective Hamiltonian separately by Taylor-Frech\'{e}t  expansion around a reference
field configuration $t$, so that $s = t + \phi$. The Hamiltonian for $\phi$ is
then 
\begin{eqnarray}\label{eq:HeffTaylor}
H[\phi] &= & H_0  - j^\dagger \phi + 
\frac{1}{2}\,\phi^\dagger D^{-1}\, \phi + \nonumber\\
&&
 \sum_{n=3}^{\infty} \frac{1}{n!} \,\Lambda^{(n)}_{x_1\ldots x_n}\, \phi_{x_1}\cdots \phi_{x_n}, \;\mbox{with}\nonumber\\
H_0 &=& H[t] =  -\log \int \!\!dp\, e^{-H_p[t]-E_p},\nonumber\\
j_x &=& - \left. \frac{\delta H[s]}{\delta s_x} \right|_{s=t} = 
- \left\langle  \frac{\delta H_p[s]}{\delta s_x}\right\rangle_{(p|d, s=t)},\\
D^{-1}_{x\, y} &=&  \left. \frac{\delta^2 H[s]}{\delta s_x\,\delta s_y} \right|_{s=t}  \nonumber\\
&=&  \left\langle \frac{\delta^2 H_p[s]}{\delta s_x\,\delta s_y}  
- \frac{\delta H_p[s]}{\delta s_x} \, \frac{\delta H_p[s]}{\delta s_y}
\right\rangle_{(p|d, s=t)} \!\!\!\!\!\!\!\!\!\!\!\!\!\!\!\!\!\!\!\!\!
+j_x\,j_y,\; \mbox{and} \nonumber\\
\Lambda^{(n)}_{x_1\ldots x_n} &= & \frac{1}{n!} \sum_{\pi \epsilon \mathcal{P}}  \left. \frac{\delta^n \,H[s] }{\delta s(x_{\pi(1)})\cdots \delta s(x_{\pi(n)})}  \right|_{s=t}. \nonumber
\end{eqnarray}

Here, $\langle \ldots \rangle_{(p|d, s)} \equiv \int \! dp\, \ldots\,P(p|d,s)$ provides expectation values with
respect to the parameter $p$ given the data $d$ and the signal  $s$. Repeated
coordinate indices are thought to be integrated over. The interaction coefficients $\Lambda^{(n)}_{x_1\ldots x_n}$ 
are symmetrized by averaging over all possible permutations $\pi$ from the space of permutations $\mathcal{P}$. In general,
$D^{-1}_{x\, y}$ needs to be symmetrized, too, but we have left out the
symmetrization in the above equation for convenience, since in the cases we
consider $D^{-1}_{x\, y}$ is already symmetric.

In case the expansion was around $t=0$, then 
\begin{eqnarray}
\label{eq:HOeffs=0}
H_0 &=& -\log \int \!\! dp \, e^{-H_{0,p} - E_p},\nonumber\\
\label{eq:jeffs=0}
j &=& \langle j_{p} \rangle_{(p|d, s=0)}, \nonumber\\
\label{eq:Deffs=0}
D^{-1} &=&  \langle D_{p}^{-1}  - j_{p} \, j_{p}^\dagger \rangle_{(p|d, s=0)} +j\,j^\dagger, \;\mbox{and}\\
\label{eq:L3effs=0}
\Lambda^{(3)} &=& \langle\Lambda^{(3)}_p  + 3\, D_{p}^{-1} \otimes j_p  - j_p j_p j_p\rangle_{(p|d, s=0)}\nonumber\\
&& - 3\, D^{-1} \otimes j + jjj\nonumber\\
\label{eq:L4effs=0}
\Lambda^{(4)} &=& \langle\Lambda^{(4)}_p  +   4\,  \Lambda^{(3)}_p  \otimes j_p   -3\, D_{p}^{-1} \otimes  D_{p}^{-1} +6\, D_{p}^{-1} \otimes j_p j_p^\dagger \nonumber\\
&& - j_p j_p j_p j_p\rangle_{(p|d, s=0)}-  4\,  \Lambda^{(3)}  \otimes j   +3\, D^{-1} \otimes  D^{-1}\nonumber\\
&&   -6\, D^{-1} \otimes j j^\dagger     + j j j j, \;\ldots \nonumber
\end{eqnarray}

Here, an implicit tensor notation was used, with e.g. $(j\,j\,j)_{xyz} \equiv
j_x\, j_y\, j_z$ and we defined the symmetrized tensor
product $(A\otimes j)_{x_1 x_2 x_3} \equiv \frac{1}{3!} \sum_{\pi \in
  \mathcal{P}} A(x_{\pi(1)}, x_{\pi(2)}) \, j(x_{\pi(3)})$. For higher rank tensors, 
the symmetrized tensor product is defined in an analogous way.

\section{Signal spectrum uncertainty}\label{sec:ssu}
\subsection{Spectrum parameterization}

Our example application of IFT with parameter uncertainties is the reconstruction of a Gaussian signal with unknown variance, which we introduce now.

The signal covariance $(S_p)_{x\,y} = \langle s_x \overline{s_y}\rangle_{(s|p)}$ may exhibit any dependence on the spatial coordinates as long as the matrix is symmetric and positive definite.  In the cosmological relevant case of translationally and rotationally invariant signal statistics, the signal covariance is fully characterized by its power spectrum. This means, there is an orthonormal basis $O$ of the signal Hilbert space which diagonalizes $S_p$:
\begin{equation}
 (O\,S_p O^\dagger)_{k\,q} \equiv O_{k\, x} (S_p)_{x\, y} \overline{O_{q\,y}} = 1_{k\,q}\, P_{S_p}(k),
\end{equation}
with $1_{k\,q}$ the identity in the transformed basis, $P_{S_p}(k)$ the power-spectrum, and using Einstein sum convention. In case we are dealing with a signal over a $d$-dimensional Cartesian space, $O_{k\, x} = \exp(i\,k\,x)$ is simply a Fourier transformation and the Fourier space identity is $1_{k\,q} = (2\,\pi)^d \, \delta(k-q)$, provided the scalar product in Fourier space is adopted as 
$a^\dagger b = (2\,\pi)^{-d}\, \int dk \,\overline{a(k)}\, b(k) $. However, since the theory should also be 
applicable in curved spaces like the sphere, or even in spaces without translational invariance, we formulate it in an 
abstract way and just assume that the basis $O$ diagonalizes the signal covariance, which is always possible.

In general, the signal covariance $S_p$ may also exhibit any dependence on the unknown parameter $p$ of the problem, as the power spectrum in cosmology is a complicated function of the cosmological parameters. However, in order not to depend on a specific model, we model the power spectrum as being a linear combination of a number of positive basis functions $f_i(k)$ with disjunct supports (the spectral bands) with respect to the basis $O_{k\,x}$, so that
\begin{equation}
P_{S_p}(k) = \sum_i p_i f_i(k)
\end{equation}
is positive for all $k$ (all coefficients of $p=(p_i)_i$ are positive and the spectral bands cover the full $k$-space domain).
We define 
\begin{equation}
 (S_i)_{xy} = \overline{O_{k\,x}} \, f_i(k) \, O_{k\, y} 
\end{equation}
and therefore have
\begin{equation}
 S_p = \sum_i p_i S_i.
\end{equation}
Since we also need the inverse of the covariance matrix we further define
\begin{equation}
 g_i(k) = \left\{ 
\begin{array}{ll}
1/f_i(k) & f_i(k)>0 \\ 
0 & f_i(k) =0
                  \end{array}
\right.
\end{equation}
and the pseudo-inverse of the band-variances,
\begin{equation}
 S_i^{-1} = \overline{O_{k\,x}}\, g_i(k)\, O_{k\, y},
\end{equation}
so that 
\begin{equation}
S_p^{-1} = \sum_i p_i^{-1} S_i^{-1},
\end{equation}
is the inverse of $S_p$, as one can easily verify.

\subsection{Spectral prior and joint Hamiltonian}

For definiteness, we assume that the individual signal-band amplitudes $p_i$ have independent prior distributions,
\begin{equation}\label{eq:pall-prior}
 P(p) = \prod_i P(p_i), 
\end{equation}
with the individual priors being given by inverse Gamma distributions, which are power-laws with 
exponential low-amplitude cutoff at $q_{i}$ :
\begin{equation}
\label{eq:p-prior}
P(p_i) = \frac{1}{q_i\,\Gamma(\alpha_i -1)} \,\left(\frac{p_i}{q_i}\right)^{-\alpha_i} \, 
\exp \left(-\frac{q_i}{p_i}\right).
\end{equation}
For $\alpha_i \gg 1$ this is an informative prior, where  $q_i/\alpha_i$ determines the preferred value.
A non-informative prior would be given by Jeffreys prior with $\alpha_i = 1$ and $q_i=0$.%
\footnote{Since this would result in an improperly normalized prior, we understand this as $\alpha_i= 1+ \epsilon $, $q_i = \epsilon$, and $\lim_{\epsilon \rightarrow 0}$ at the end of the calculation.
We note, that this limit might not exist, or that it provides trivial results. I.e. we will find in Sec. \ref{sec:lognormal prior}
that in this limit the signal reconstructed with the full field theory turns out to be zero and the data is assumed to be purely made of noise. 
Thus the improper Jeffreys prior is actually inappropriate for the full problem, although interesting.}

The joint Hamiltonian is therefore
\begin{equation}\label{eq:joint Hamiltonian of our problem}
 H[s,p] = H^\G_p[s] + E(p)
\end{equation}
with the parameter prior energy
\begin{equation}
 E(p) = \sum_i \left[\frac{q_i}{p_i} + \alpha_i \log\left(\frac{p_i}{q_i}\right) + \log(q_i\,\Gamma(\alpha_i -1))\right].
\end{equation}

\subsection{Generic filter formula}

In the following, we derive five approximate filters for this problem. 
It will turn out that they can all be cast into a single set of determining equations, with different coefficients. 
This generic filter formula should be presented first, before we discuss the individual approaches.

All of the derived filters can be expressed as Wiener filters for some specific spectrum $S_{p^*} = \sum_i p_i^* S_i$, with different spectral parameters $p^*$. The signal map and the spectrum assumed for its construction have to be calculated self-consistently from 
\begin{eqnarray}\label{eq:characteristic equation}\label{eq:general-spectrum}
 m_{p^*} &=& D_{p^*}\, j,\;\mbox{and}\\
p_i^* &=& \frac{1}{\gamma_i+\varepsilon_i} \left( q_i +  \frac{1}{2} \mathrm{Tr}[(m_{p^*} m_{p^*}^\dagger + \delta_i\, D_{p^*})\, S_i^{-1}]\right),\nonumber
\end{eqnarray}
for example by simply iterating these two equations. 

Here,  the filter-specific parameters are $\varepsilon_i$, $\delta_i$, and $\gamma_i = \alpha_i - 1 + \varrho_i/2$, where $\varrho_i = \mathrm{Tr}[S_i^{-1}S_i]$ is the number of degrees of freedom of the $i$th spectral band. 
In order to simplify notation, we drop in the following the $*$ from $p^*$, assuming that the context makes it clear wether we talk about the unknown parameter $p$ or a parameter choice $p^*$ for a specific filter. 

In order to develop a filter for our signal, we have to decide according to which principle the signal or the power spectrum used in the Wiener filtering is determined. 
In the space of all possibilities for the signal and its power spectrum the joined probability function $P(s,p|d)$ has to be asked. 
There are different hyperplanes in this space along which this function can be cut, marginalized, and maximized. 
The ultimate answer of the PURE approach will come from marginalizing $p$ and calculating the signal mean.
However, first we want to establish more traditional signal estimators, using largely the MAP principle along different cuts through the joint signal and spectral parameter space.

In case a Jeffreys prior is adopted ($q_i=0$ and $\alpha_i=1$) it will turn out that the trivial filter $m(d)=0$  would be the preferred solution in all cases. 
However, since Jeffreys prior is an improper prior which is convenient to represents the class of very broad, but proper priors, we should not hesitate to remove the trivial filter solution by hand. 
Otherwise we would need to enter the discussion about an appropriate informative prior, which we like to avoid for simplicity.
This can not be decided generically, but only for any concrete inference problems individually.

\begin{figure}[tb]
 \centering
 \begin{center}
 \includegraphics[bb= 30 320 570 740,width=0.8\columnwidth]{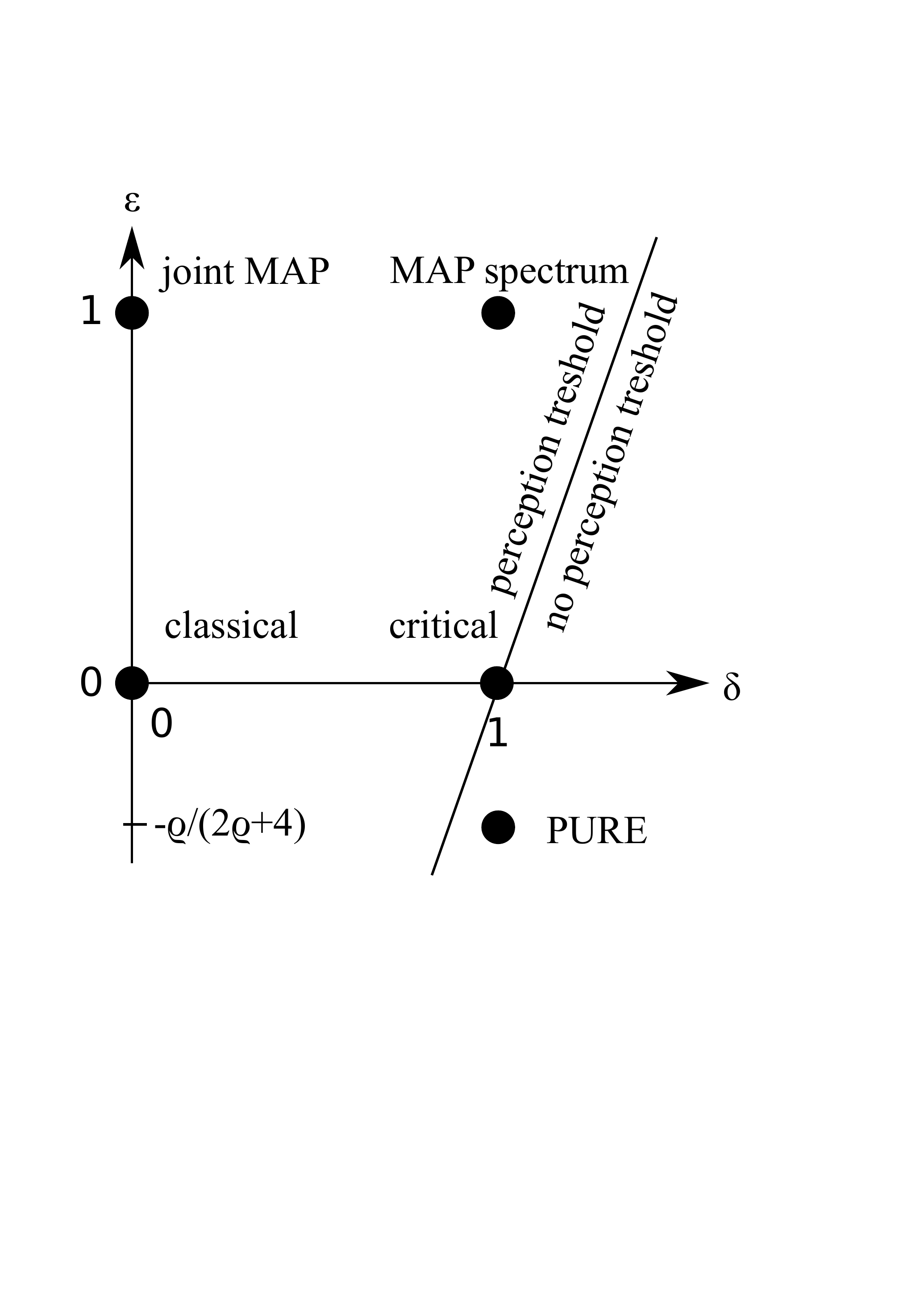}
\end{center}
 \caption{Parameter $\delta_i$ and $\varepsilon_i$ of the five different filters for Jeffreys prior in the representation of the generic filter formula Eq.\ \ref{eq:characteristic equation}. 
The parameter of the displayed filter are derived in the following sections: the critical filter in Sec.\
\ref{sec:critical filter}, the classical filter in \ref{sec:classical filter}, the joint MAP filter in Sec.\ \ref{sec:jMAP}, the MAP spectrum filter in Sec.\ \ref{sec:MAPspecFilter}, and the PURE filter in Sec. \ref{sec:PUREprojected}.
The critical line between filter with and without perception threshold as given by Eq.\ \ref{eq:critical line} is also shown. 
}
 \label{fig:diagram}
\end{figure}

The parameters of the filters described in Sec. \ref{sec:derivedfilters} and 
derived in the next few subsections are summarized in Fig.\ \ref{fig:diagram}.

\subsection{The critical filter}\label{sec:critical filter}

Our first filter can be understood without any reference to statistical inference and is along the lines of 
the well known Karhunen-Lo\`eve (KL, \citep{1947KarhunenK,1978LoeveM, 2002MNRAS.335..887T}) and  Feldman-Kaiser-Peacock (FKP,   \citep{1994ApJ...426...23F}) estimators for power spectra.
The Wiener filter map $m_p = D_p\, j$ (with $D_p = (S_p^{-1} + R^\dagger N^{-1} R)^{-1}$ and $j =  R^\dagger N^{-1} d$) of a data realization of a Gaussian random signal with a known covariance $S_p$ will have on average the covariance
\begin{equation}
 \langle m_p \, m_p^\dagger \rangle_{(d,s|p)} = S_p - D_p,
\end{equation}
as one can verify with a short calculation.\footnote{%
Using the abbreviation $M = R^\dagger N^{-1} R$ we write
$\langle m_p \, m_p^\dagger \rangle_{(d,s|p)} = 
D_p\,  \langle j \, j^\dagger \rangle_{(d,s|p)} \, D_p = 
D_p R^\dagger N^{-1} ( R\, S_p R + N) \, N^{-1} R \, D_p = 
D_p (M\, S_p M + M) \, D_p = 
D_p M \, (1+ S_p M) \, (1+ S_p M)^{-1} S_p  =
D_p M \,S_p = 
S_p (1 + M\, S_p)^{-1}  (1+ M\, S_p -1) =
S_p - D_p$.}
The propagator on the rhs just accounts for the power lost in measurement and filtering. 
Now we assume that our data and our Wiener filter map are so rich or typical that this equation also holds for our individual data realization. 
Thus we drop the expectation angles, apply $\mathrm{Tr}[\times\,S_i^{-1}]$, and get the critical filter recipe
in the form of Eq. \ref{eq:characteristic equation} with parameters $\delta_i = 1$, $\varepsilon_i =0$, $\alpha_i = 1$, and $q_i =0$. 
The last two parameters are characteristic for Jeffreys prior, which we obviously have assumed implicitely, since no prior information on the spectrum, or even its magnitude on a logarithmic scale, has entered the critical filter scheme. 

The name \textit{critical filter} should become clear in Sec.\ \ref{sec:perception threshold}. 
There, we show that at least in cases where the different spectral parameters are independent of each other, the different filters can be cast into two classes, such with and such without perception threshold. 
The critical filter marks the demarcation line between these phases.

The critical filter has recently been applied successfully by \cite{2010arXiv1008.1246O} to reconstruct an all sky map of the galactic Faraday depth from sparse and noisy measurements.

\subsection{Joint MAP filter}\label{sec:jMAP}

Extremizing the joint Hamiltonian, Eq.\ \ref{eq:joint Hamiltonian of our problem}, with respect to $p$ and $s$ yields the joint MAP filter parameters $(\delta_i, \varepsilon_i)= (0,1)$. We note, that if we extremize with respect to the log-spectral amplitudes $\tau_i= \log p_i$, the parameters $(\delta_i, \varepsilon_i)= (0,0)$ would have resulted due to the effect of the Jacobian of the prior transformation. This latter  filter is identical to the classical one derived below in Sec. \ref{sec:classical filter}.

\begin{figure}[t]
 \centering
 \begin{center}
\includegraphics[bb=180 160 650 480,width=\columnwidth]{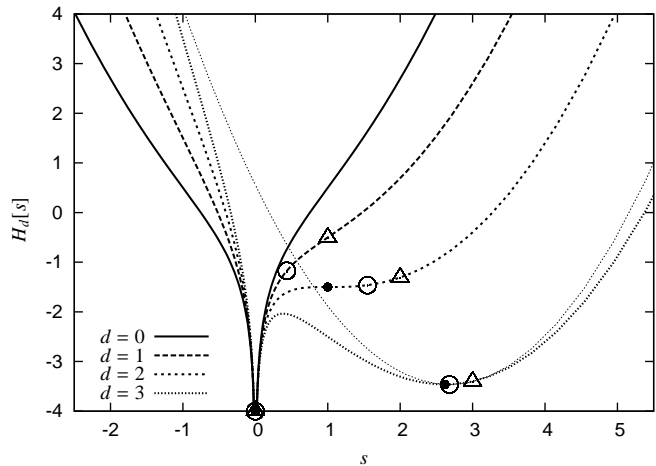}
\end{center}
 \caption{The effective signal Hamiltonian, Eq.\ \ref{eq:MAPmapHamilton}, without the normalization constant $H_0$ in case of Jeffreys prior and for a single, independent signal $s=s_i$ and data point $d=d_i$. The parameters are $R_{ij}=N_{ij}=\delta_{ij}$ and $S_{ij} = p_i\, \delta_{ij}$. The different curves show the Hamiltonian  for representative data values. The triangle symbols mark the results of the inverse response estimator $m_\mathrm{ir} = R^{-1} d$ on the corresponding curves. The large open and small filled circles mark the re\-norma\-lized and classical map estimator results, respectively. The existence of a classical perception threshold can be seen: for $-2<d<2$, the classical map is exact zero since no non-trivial stationary point of the Hamiltonian exists. The thin dotted line shows the renormalized Hamiltonian for the case $d=3$, as provided by $\frac{1}{2}(s-m_p)^\dagger D_p^{-1} (s-m_p)$.}
 \label{fig:H(s)}
\end{figure}

\subsection{MAP spectrum filter}
\label{sec:MAPspecFilter}

Marginalizing the joint Hamiltonian Eq.\ \ref{eq:joint Hamiltonian of our problem} over the signal space provides the spectrum Hamiltonian
\begin{eqnarray}\label{eq:MAPspecHamilton}
 H(p) &=& - \log(P(d,p)) = - \log(P(d|p)\, P(p))\nonumber\\
&=& \frac{1}{2}\log| 1+ Q_p| -   \frac{1}{2} j^\dagger D_p\, j + H_0'\nonumber\\
&+& \sum_i \left(\frac{q_i}{p_i} + \alpha_i \log \left( \frac{p_i}{q_i}\right) \right),\; \mathrm{with}\\
D_p &=& (S_p^{-1} + M)^{-1},\; Q_p = S_p \, M,\; \mathrm{and}\nonumber\\
H_0' &=& \frac{1}{2}\log|N| + \frac{1}{2} d^\dagger N\, d +  \sum_i \, \log( q_i\, \Gamma(\alpha_i -1)).\nonumber
\end{eqnarray}
Here we used Eq.\ \ref{eq:ZdfreeTheory} for $P(d|p)$. A data-space view on this likelihood is given in Appendix \ref{sec:signal covariance like}.
Extremizing $H(p)$ with respect to $p_i$ and sorting for terms linear in it provides the MAP-spectrum parameter $(\delta_i,\varepsilon_i ) = (1,1)$. 

If we extremize with respect to the parameters $\tau_i = \log p_i$, we get $(\delta_i, \varepsilon_i)= (1,0)$, the parameters of the critical filter. 
Thus, the critical filter can be regarded as the one resulting from a MAP spectrum estimation on a logarithmic scale. 
Note that MAP estimators are sensitive to the coordinate system in which parameters are expressed.

\subsection{Classical map estimator}\label{sec:classical filter}
The effective, parameter marginalized signal Hamiltonian (Eq.\ \ref{eq:effectiveHamiltonianGeneral})
can be calculated analytically:\footnote{The term $|S_i|$ has to be read as the determinant within the non-zero subspace of $S_i$.}
\begin{eqnarray}\label{eq:MAPmapHamilton}
 H[s] &=& \frac{1}{2} \, s^\dagger M\, s - j^\dagger s + \sum_i  \gamma_i\, \log\left(q_i+ \frac{1}{2} \, s^\dagger S_i^{-1} s\right)\nonumber\\
&+& H_0
,\;\mbox{with}
\nonumber\\
H_{0} &=& \frac{1}{2} d^\dagger N^{-1} d  + \frac{1}{2} \log\left(|2\,\pi\,N|\right)\nonumber\\
&-& \log\left( \prod_i \frac{\Gamma(\gamma_i)\, q_i^{\alpha_i -1}}{\Gamma(\alpha_i -1)\,|2\,\pi\,S_i|^{\frac{1}{2}}}\right).
\end{eqnarray}

The classical mapping equation results from extremizing this Hamiltonian and is provided by Eq.\ \ref{eq:general-spectrum} for $(\delta_i, \varepsilon_i) = (0,0)$.
This can be regarded as a poor man's critical filter, since only the power in the map is used to determine the signal covariance, and no correction for the power lost in the filtering is applied.
In case of a single independent data and signal point, the effective Hamiltonian is an one dimensional function in signal space and is shown in Fig.\ \ref{fig:H(s)}.

\section{Uncertainty renormalization flow}\label{sec:uncertainty renormalization flow}
\subsection{General remarks}

Although the MAP methods often provide acceptable signal estimators, they are not optimal in an $\mathcal{L}^2$-error norm sense. 
In case of a skewed posterior, such reconstructions are suboptimal. 
Our goal is to calculate moments of the signal field averaged over the effective posterior, as e.g. $\langle s \rangle_{(s|d)}$ given by Eq. \ref{eq:m_eff}, since those optimize the $\mathcal{L}^2$-error. 
For this we might construct the effective Hamiltonian exactly or in terms of a Taylor expansion as in Eqs. \ref{eq:HeffTaylor} and  \ref{eq:Deffs=0}.

Such an expansion of the effective Hamiltonian around a reference field is expected to work best when the para\-meter prior is well localized around a specific value. 
The effective Hamiltonian will then be close to the original, parameter-dependent one for this parameter value. 
In case the original theory was free, the effective theory will have only small interaction terms. 
Diagrammatic expansions can then be conducted and truncated at low order. 

Unfortunately, in many practical applications, the uncertainties of the parameters are substantial, and not described by a well localized prior. In this case it might be possible to construct  the effective Hamiltonian by repeatedly adding smaller portions of parameter-uncertainty, with each uncertainty dose so small that the resulting Hamiltonian has only weak interactions, which can be re-absorbed into renormalized, effective propagator and data source terms. The accumulated uncertainty can thereby become large and equal to the required amount of entropy for the unknown parameter of the theory. In the following we will explain the basics of this uncertainty renormalization flow. 

\subsection{Parameter uncertainty renormalization}

A broad prior for a parameter $p$ may be decomposed into a number $N$ of narrow and mutually independent priors for some auxiliary variables $\tau_j$ (with $j \in \{1,...,N\}$):
\begin{equation}
 P(p) = \left( \prod_{j=1}^{N}\, \int \! d\tau_j \, P(\tau_j)\right) \delta(p - \sum_{j=0}^{N} \tau_j).
\end{equation}
We have chosen here the parameter to be the sum of the auxiliary variables for definiteness and simplicity, but other relations can be worked out in a similar way or be mapped onto this case. Also the mutual independence of the auxiliary variable is mostly a technical convenience and not a strict requirement. 
Note, that we have included a starting parameter value of $\tau_0$ into the sum. Since it would be convenient to identify this with the prior expectation value $ \langle p \rangle_{(p)} $ throughout the full renormalization procedure we require
\begin{equation}\label{eq:unbiased auxiliary priors}
\langle \tau_j \rangle_{(\tau_j)} = \delta_{j \, 0} \, p_0,
\end{equation}
with $p_0 = \langle p \rangle_{(p)}$.
We further introduce the $l$-th parameter residual as $r_l = \tau_0 + \sum_{j=l+1}^{N} \tau_j$, so that $r_0 = \sum_{j=0}^{N} \tau_j = p$ and $r_N = \tau_0 = p_0$.
The effective Hamiltonian can now be expressed as
\begin{eqnarray}
 e^{-H[s]} &=& \int \!dp \, P(p) \, e^{-H_p[s]} \nonumber\\
&=&   \int \!dp \, \left( \prod_{j=1}^{N}\, \int \! d\tau_j \, P(\tau_j)\right) \delta(p - \sum_{j=0}^{N} \tau_j)\, e^{-H_p[s]} \nonumber\\
&=&   \left( \prod_{j=1}^{N}\, \int \! d\tau_j \, P(\tau_j)\right) \,  e^{-H_{r_0·}[s]} \nonumber\\
&=&  \left( \prod_{j=2}^{N}\, \int \! d\tau_j \, P(\tau_j)\right) \, \underbrace{ \int \! d\tau_1 \, P(\tau_1)\,  e^{-H_{\tau_1+r_1·}[s]}}_{\equiv e^{-H_{r_1·}^{(1)}[s]} }  \nonumber\\
&=&  \left( \prod_{j=3}^{N}\, \int \! d\tau_j \, P(\tau_j)\right) \,  \underbrace{\int \! d\tau_2 \, P(\tau_2)\,  e^{-H_{\tau_2+r_2·}^{(1)}[s]} }_{\equiv e^{-H_{r_2·}^{(2)}[s]} } \nonumber\\
&=& ... \nonumber\\
&=&  \int \! d\tau_N \, P(\tau_N) \,   e^{-H_{\tau_N·+p_0}^{(N-1)}[s]} = e^{-H_{r_N=p_0}^{(N)}[s]},
\end{eqnarray}
where $H_{p_0}^{(N)}[s] =  H[s]$.
This means that a series of effective Hamiltonians with increasing accumulated parameter uncertainty is defined, and an uncertainty adding operator:
\begin{equation}\label{eq:Hrecast}
 H_{r_n}^{(n)} \mapsto  H_{r_{n+1}}^{(n+1)}  \equiv - \log  \int \! d\tau_{n+1} \, P(\tau_{n+1})\, e^{-H_{r_{n+1}+\tau_{n+1}}^{(n)}}.
\end{equation}
Note that $H_{r_0}^{(0)}=  H_{p=r_0}$ and $H_{r_N}^{(N)} = H$.
This uncertainty renormalization can be done using Eq. \ref{eq:HeffTaylor} if it is not possible to do it analytically.
To each Hamiltonian a time-like variable $t$ can be assigned, which measures the amount of uncertainty accumulated so far. A suitable variable is the accumulated uncertainty dispersion,
\begin{equation}
 t_n = \sum_{j=1}^{n} \sigma_{\tau_j}^2 = \sum_{j=1}^{n} \left( \langle \tau_j^2 \rangle_{(\tau_j)} -  \langle \tau_j \rangle_{(\tau_j)}^2\right) 
=  \sum_{j=1}^{n} \langle \tau_j^2 \rangle_{(\tau_j)} ,
\end{equation}
where we used Eq.\ \ref{eq:unbiased auxiliary priors}. In case all auxiliary variables have the same prior, we find $t_n = n \, t_1 = \frac{n}{N}\, \langle (p-p_0)^2 \rangle_{(p)}$. 

At each time-step a renormalization of the Hamiltonian can be done, in which it is cast back into the structure it had before, 
e.g. in our example of reconstruction with unknown power spectrum the free Hamiltonian of Eq. \ref{eq:freeHamiltonian},
just with modified coefficients (propagator, source and interaction terms). 

In our example the recast Hamiltonian is free, which implies that we are constructing a Gaussian approximation of the parameter marginalized signal posterior to be used for inference. 
It is shown in  \cite{2010PhRvE..82e1112E} that the chosen Gaussian seems to be optimal in an information theoretical and thermodynamical sense. It maximizes the cross information with the correct effective posterior.

A renormalization flow can further be established by letting the individual time-steps of size $t_1$ become infinitesimally small, however, their number $N$ infinitely large, while keeping the total added uncertainty constant, $t= N\, t_1$. 
The result are the renormalization flow equations for the coefficients of the Hamiltonian. 
The actual form of these equations depends on the Hamiltonian and is much simpler if the Hamiltonian has less interactions.

Therefore, even a free Hamiltonian as in Eq.\ \ref{eq:freeHamiltonian}
should be further simplified by suppressing the linear term $j^{\dagger}_p s $ as far as possible. This is done  for the value of $p=p_0$, which is our starting point in parameter space, by changing to a new field variable $\phi = s -m_0$ with $ m_0 \equiv D_{0} \, j_{0}$, $D_{0} \equiv D_{p_0} $, and $j_{0}\equiv j_{p_0}$. The Hamiltonian reads now
\begin{eqnarray}\label{eq:defHamiltonianwithPara1}
H'_p[\phi] &=& \frac{1}{2} \phi^\dagger D_p^{-1} \phi -  {j'_p}^\dagger \phi + H'_{0,p},\;\mbox{with}\nonumber\\
\phi &=& s - m_0 , \nonumber\\
j'_p &=& j_p - D_p^{-1}\,m_0 = j_p-D_p^{-1}\,D_0\, j_0,\;\mbox{and}\\
H'_{0,p} &=&  H_{0,p}  +  \frac{1}{2} m_0^{\dagger} D_p^{-1} m_0 - j_p^\dagger m_0\nonumber
\end{eqnarray}
and is especially simple for $p=p_0$, since then $j'_0  = 0$. Now, the effective Hamiltonian is calculated and expanded according to the recipe in Sec.\ \ref{sec:effectiveH} for a parameter prior well localized on $p=p_0$. The localization of the prior is typically characterized by a small parameter $\delta t= \sigma_\tau^2 $, which also appears as a pre-factor of the various coefficients of the effective Hamiltonian. 

\subsection{4$^\mathrm{th}$ order interactions}

In order to perform the renormalization step, 
the recasting of the uncertainty marginalized Hamiltonian in Eq. \ref{eq:Hrecast} into its original form,
let us be a bit more specific about the effective Hamiltonian for definiteness. By virtue of our foresight on the calculations in Sec.\ \ref{sec:lognormal prior} we assume that up to linear order in $\delta t$ the effective Hamiltonian is given by 
\begin{eqnarray}
H'[\phi] &= &  H_0' + \Lambda^{(1)\dagger} \phi +\frac{1}{2}\,\phi^\dagger \left( {D}_0^{-1} +\Lambda^{(2)} \right) \, \phi \\
&& + \frac{1}{3!} \,\Lambda^{(3)}\, [\phi, \phi, \phi] + \frac{1}{4!} \,\Lambda^{(4)}\, [\phi, \phi, \phi, \phi]+ \mathcal{O}(\delta t^2),\nonumber
\end{eqnarray}
with $ \Lambda^{(1)}$, $\Lambda^{(2)} $,  $\Lambda^{(3)} $, and  $\Lambda^{(4)} $ being of order 
$\mathcal{O}(\delta t)$. 
Here, Eq. \ref{eq:HeffTaylor} or \ref{eq:Deffs=0} might have been used.

The corrections can be expected to be small of $\mathcal{O}(\delta t)$, since our originally free Hamiltonian, Eq.\ \ref{eq:defHamiltonianwithPara1}, should be recovered in the limit of vanishing parameter uncertainty, $\delta t \rightarrow 0$.
All higher order interaction terms are of higher order in $\delta t$ and therefore ignored in the following. For our later convenience we introduce 
\begin{equation}
 \lambda^{(n)} =\lim_{\delta t \rightarrow 0} \frac{\Lambda^{(n)}}{\delta t}.
\end{equation}

Now, we can renormalize by absorbing all diagrams of order  $ \mathcal{O}(\delta t)$ into renormalized propagator and source terms,
in order to obtain a free Hamiltonian.
Since $j'$ in Eq.\ \ref{eq:defHamiltonianwithPara1} is already of order $\delta t$ and only the three- and four-leg vertices have contributions of order $\mathcal{O}(\delta t)$, only uncertainty loop corrections have to be taken into account. We can therefore define the renormalized data-source vertex of the effective $\phi$-theory,
\begin{eqnarray}
j_\# &=& 
\includegraphics[width=0.23\fgwidth]{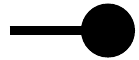}+
\includegraphics[width=0.8\fgwidth]{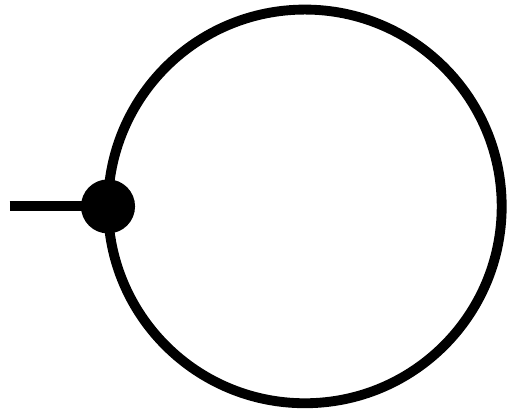}+\mathcal{O}(\delta t^2)\nonumber\\
&=& -\Lambda^{(1)}  - \frac{1}{2}\, \Lambda^{(3)} [\cdot, D]\\
&=& -\Lambda^{(1)}_x - \frac{1}{2}\, \Lambda^{(3)}_{x y z}  D_{y z},\nonumber
\end{eqnarray}
which takes the dominant uncertainty-loop correction into account. We dropped the subscript at $D_0$ and use the Feynman rules provided in \cite{2009PhRvD..80j5005E}.
The renormalized propagator up to linear order in $\delta t$ is
\begin{eqnarray}
D_\# &=& \includegraphics[width=\fgwidth]{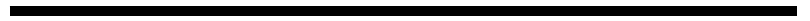}+\includegraphics[width=\fgwidth]{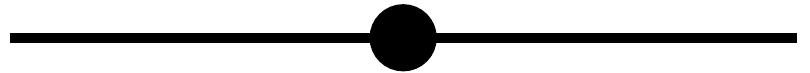}+\includegraphics[width=\fgwidth]{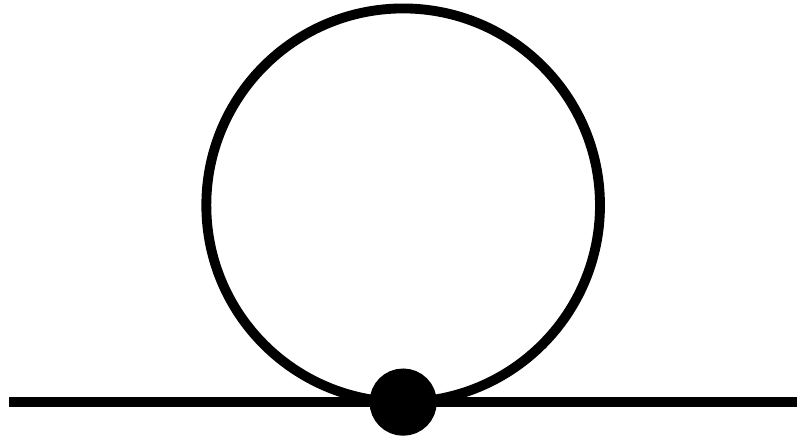}+\mathcal{O}(\delta t^2)\nonumber\\
&=& D - D\, \Lambda^{(2)} \, D  - \frac{1}{2}\,D\, \Lambda^{(4)}[\cdot, D, \cdot]\, D\\
&=& D_{x y} - D_{x z} \, \Lambda^{(2)}_{z z'} \, D_{z' y}  - \frac{1}{2}\, D_{x x'} \, \Lambda^{(4)}_{x' z z' y'}\,D_{z z'}\,D_{y' y}.\nonumber
\end{eqnarray}
The inverse renormalized propagator up to first order is
\begin{eqnarray}
D_\#^{-1} &=& D^{-1} + \Lambda^{(2)}  + \frac{1}{2} \, \Lambda^{(4)} [\cdot, D, \cdot]\,.
\end{eqnarray}

These coefficients now define a renormalized effective Hamiltonian, 
$H'_\#[\phi] = \frac{1}{2} \phi^\dagger D_\#^{-1} \phi - j_\#^\dagger \phi + H_{\#0}$,
which belongs to a free theory, and is similar to $H'[\phi]$, in that it has the same mean and uncertainty dispersion by construction. Higher order uncertainty correlations differ certainly, due to the approximation of the renormalization step. In contrast to the original Hamiltonian $H_{p_0}[s]$ in Eq.\ \ref{eq:freeHamiltonian}, which was also free, the renormalized Hamiltonian has some amount of parameter uncertainty corrections included.

Now, the original field $s= m_0+\phi $ can be restored, leading to a free Hamiltonian with
\begin{eqnarray}
D_{t+\delta t} &=& D_\#,\;\mbox{and} \\
j_{t+\delta t} &=&  j_\# +D_\#^{-1} m_t,\nonumber
\end{eqnarray}
where the subscript $t+\delta t$ indicates that the parameter uncertainty is increased by $\delta t$ from its original value of $t$. Since $\delta t$ can be made arbitrarily small, a system of differential equations can be derived,
\begin{eqnarray}\label{eq:Ddot}
\frac{dD_t}{dt} &=& \lim_{\delta t \rightarrow 0} \frac{D_{t+\delta t}-D_{t}}{\delta t}\nonumber\\
&=&  - D_t\, \lambda^{(2)} \, D_t  - \frac{1}{2}\,D_t\, \lambda^{(4)}[\cdot, D_t, \cdot]\, D_t
,\;\mbox{and} \nonumber\\
\frac{dj_t}{dt} &=& \lim_{\delta t \rightarrow 0} \frac{j_{t+\delta t}-j_{t}}{\delta t}\\
&=&  -\lambda^{(1)}  + \lambda^{(2)} D_t\, j_t - \frac{1}{2}\, \lambda^{(3)} [\cdot, D_t]\nonumber\\
&+& \frac{1}{2} \, \lambda^{(4)} [\cdot, D_t,  D_t\, j_t ],\nonumber
\end{eqnarray}
which form the uncertainty renormalization flow equations. The pseudo-time $t$ measures the accumulated dispersion of the resulting prior probability. These equations can be transformed into the more compact form
\begin{eqnarray}
\frac{dD_t^{-1}}{dt} &=&    \lambda^{(2)}   + \frac{1}{2}\, \lambda^{(4)}[\cdot, D_t, \cdot]
,\;\mbox{and} \\
\frac{dm_t}{dt} &=&  -D_t\,\lambda^{(1)}    - \frac{1}{2}\, D_t\, \lambda^{(3)} [\cdot, D_t],\nonumber
\label{eq:mdot}
\end{eqnarray}
where $m_t = D_t \, j_t$. 

The renormalization equations so far are evolution equations for operators. 
If they should become ordinary partial differential equations, e.g. in our case in terms of spectral parameters, some sort of closure is required. 
This should ensure that the renormalized Hamiltonian gets its original structure, so that it is clear which terms are affected by the parameter uncertainty adding operation. 
Ideally, the change in the Hamiltonian can be mapped onto changes of effective parameter values.

After the repeated adding of small amounts of parameter uncertainty, the resulting effective parameter prior distribution can be expected to be a Gaussian, due to the central limit theorem of statistics, 
\begin{equation}
 P(p) = \G(p-p_0, t).
\end{equation}

\section{Signal reconstruction with PURE}\label{sec:applyingPURE}
\subsection{Lognormal spectral prior}\label{sec:lognormal prior}

Now we want to apply the PURE scheme to our example problem from Sect\ \ref{sec:ssu} of how to reconstruct a Gaussian signal with unknown covariance.

First, we have to express our spectral prior in a way that we can apply the PURE method developed in the previous section.
For this we need some additive auxiliary random variables into which we can decompose our (unknown) spectral amplitudes.
These variables should each have an unbiased distribution with zero mean according to Eq.\ \ref{eq:unbiased auxiliary priors}. 
For the moment, we concentrate on a single spectral parameter $p_i$ and change to the parameter variable $\tau_i = \log p_i$, which can be split up into additive auxiliary variables: $\tau_i = \sum_j {\tau_{ij}}$. 
For convenience we assume $p_{ij} = e^{\tau_{ij}}$ to be distributed according to Eq.\ \ref{eq:p-prior}, with properly chosen parameters $\alpha_{ij}$, and $q_{ij}$, as detailed in the Appendix \ref{sec:central limit theorem}. There, it is shown that 
\begin{equation}
 P(\tau_i)  \longrightarrow   \G(\tau_i,t_i)
\end{equation}
for the limit of an infinite number of auxiliary parameters, with a finite total uncertainty dispersion of $t_i =  \langle \tau_{i}^2 \rangle_{(\tau_{i})} - \langle \tau_{i} \rangle_{(\tau_{i})}^2$, as expected from the central limit theorem of statistics. The resulting statistics for $p_i = e^{\tau_{i}}$ is therefore log-normal. If we take the limit  $t_i\rightarrow \infty $ we obtain Jeffreys prior, which is flat on a logarithmic scale, and which conveniently permits us to compare the PURE filter to the others.

\subsection{Uncertainty renormalization}

In the following we assume that all spectral coefficients receive uncertainty with the same infinitesimal rate, so that the prior distributions in Eq.\ \ref{eq:p-prior} are all the same and narrowly centered on $p_i=1$, which implies
$\delta t_i  = 1/(\alpha_i -1)= \delta t$ and $q_i=\alpha_i-3/2 = \delta t^{-1} -1/2$ (see Appendix \ref{sec:central limit theorem}).

Expanding the Hamiltonian in Eq.\ \ref{eq:MAPmapHamilton} around the reference map $m=D \, j$ recovers the original free Hamiltonian, shifted to $\phi = s -m$, and perturbed by some additional interaction terms $\Lambda^{(n)} = \delta t \,\lambda^{(n)}+ \mathcal{O}(\delta t^2)$ with
\begin{eqnarray}
 \lambda^{(1)} &=& \sum_i \frac{1}{2} \left(\varrho_i +1  - p^{-1} m^\dagger S_i^{-1} m\right) S_i^{-1} p_i^{-1} m , \nonumber \\
 \lambda^{(2)} &=& \sum_i \frac{1}{2} \left( \varrho_i +1 - p_i^{-1} m^\dagger S_i^{-1} m\right)  \, 
 S_i^{-1} p_i^{-1}   \nonumber \\
&-& S_i^{-1} m m^\dagger S_i^{-1} p_i^{-2} , \nonumber\\
\lambda^{(3)} &=& \lambda^{(4)}[\cdot, \cdot,\cdot,     m] , \\
\lambda^{(4)} &=& -3  \sum_i \,p_i^{-2}\, S_i^{-1} \otimes S_i^{-1} ,  \; \mbox{and}\nonumber\\
\lambda^{(n)} &=&  0 \; \mbox{for}\; n>4.  \nonumber
\end{eqnarray}
Here we have reinserted $p_i$ in order to have variables which capture the evolution of the renormalization flow dynamics.
The renormalization flow equations are given by inserting the latter terms into two independent equations out of Eqs. \ref{eq:Ddot} - \ref{eq:mdot}:
\begin{eqnarray}
 \!\!\!\!\!\!\frac{dD^{-1}}{dt}\!\!\! &=& \!\!\!\sum_i \!\!
 \left[ \frac{1}{2}\left((1+\varrho_i)  p_i -\mathrm{Tr}[B_i]\right) S_i^{-1}  \!-\! S_i^{-1}  B_i  \right]\!\! p_i^{-2}\!\!\!\!\!,\,\,\,\,\, \label{eq:D-1dot}\nonumber\\
\frac{dj}{dt} &=& - \sum_i p_i^{-2} (m^{\dagger} S_i^{-1} m) \,  S_i^{-1} m
,  \;\; \mathrm{with} \label{eq:djdt}\\
B_i &=& (m\, m^\dagger + D)\, S_i^{-1} \;\; \mathrm{and} \;\; m = D\, j .\nonumber
\end{eqnarray}
This system of integro-differential equation represents the most accurate form of the PURE filter for this application.
It is, however, in general quite expensive to implement numerically, since it requires to follow the evolution of matrices.

\subsection{Projection onto spectral parameterization}

To simplify the PURE filter equations, we want to recast the system into the original from, which assumes $D^{-1} = (S_p^{-1}+ M)$ with $S^{-1}=\sum_i p_i^{-1}\, S_i^{-1}$. 
Thus evolution equations for the $p_i$s are needed. 
Since $\frac{d}{dt}D^{-1} = \sum_i S_i^{-1} \, \frac{d}{dt} p_i^{-1}+\frac{d}{dt}M$ contains the parameter evolution one has to specify how to split the evolution equation of the inverse propagator.

A natural way is to require all terms of the rhs of Eq.\ \ref{eq:D-1dot}, which are parallel to the inverse signal covariance bands, to contribute to their evolution, and the ones which are orthogonal, to contribute to the evolution of $M$. The part of a matrix $A$ parallel to $S_i$ is obtained by the projector
\begin{equation}
 \mathcal{P}_i \, A \equiv \frac{1}{\varrho_i}\mathrm{Tr}\left[A\, S_i \right] \, S_i^{-1}
\end{equation}
and the orthogonal part by $(1-\mathcal{P}_i)\,A$. Splitting the evolution equation this way yields
\begin{eqnarray}\label{eq:dpidt}
 \frac{dp_i}{dt} &=& \beta_i\, p_i, \;\;\mathrm{or}\;\; \frac{d\tau_i}{dt} = \beta_i, \;\;\mathrm{and} \\
\frac{dM}{dt} &=&  \sum_i \, p_i^{-2}  S_i^{-1}  \left(
\frac{1}{\varrho_i}\mathrm{Tr}\left[B_i \right] - B_i \right), \;\;\mathrm{with} \nonumber\\
\beta_i &=&  \left(\frac{1}{2}+\frac{1}{\varrho_i}\right) \mathrm{Tr} \left[  B_i \right] p_i^{-1} -  \frac{1+\varrho_i}{2}.\nonumber
\end{eqnarray}
With this, the fastest evolution is assigned to the signal strength, whereas the inverse noise term evolves on much longer time-scales for $\varrho_i \gg 1$. Actually, $M$ evolves only in directions orthogonal to all $S_i$, since 
\begin{equation}
\frac{d}{dt} (\mathcal{P}_i \,M) = 0,
\end{equation}
meaning that the power within the spectral bands of $M$ gets only reshuffled, but is conserved.
This implies that the evolution of $M$ interferes very little with the spectral evolution, since all changes to $M$ happen in directions which are projected out for $S_p$. The reverse is not true, since $M$ couples to the value of $p$. For an accurate reconstruction the evolution of $M$ needs to be followed, since it determines $D$ and thereby $m=D\, j$. However, we focus now only on the signal spectrum evolution and ignore the slow and perpendicular $M$ evolution. 

The evolution equation for $p$ and $j$ have to be solved simultaneously
as a function of $t$ up to the spectral uncertainty $t_\mathrm{max} = \langle (\log p - \log p_0)^2 \rangle_{(p)}$ of the original problem. 
This version of the PURE filter for spectrally uncertain Gaussian random signals with a lognormal spectral prior is projected onto our spectral parametrization, but not yet onto our generic filter formula.

\subsection{Jeffreys prior}

Let us see if there is a stationary asymptotic for the limit of infinite spectral uncertainty.
The resulting filter for $t \rightarrow \infty$ (which implies a Jeffreys prior) 
seems to be trivial, since $j\rightarrow 0$ and therefore $m_p\rightarrow 0$ in this limit. 

This can actually be understood intuitively. 
On the logarithmic scale $\tau_i = \log p_i$ Jeffreys prior becomes flat in $\tau_i$. 
Thus an arbitrary negative $\tau_i$ (and therefore infinitesimally small $p_i$) is as probable a priori as an arbitrary large $\tau_i$ (and therefore basically infinite large $p_i$).
However, the likelihood $P(d|p) = \int \!\mathcal{D}s\, P(d,s|p)$ discriminates clearly between those cases.

For $p\approx 0$ we expect $s\approx 0$, which means that the data must be purely noise, which has a low, but finite likelihood. 
This likelihood does not decrease significantly if $\tau \rightarrow -\infty$ and $p$ and $s$ become exactly zero, since the amount of noise stays constant. 
It has to be identical to the data in this case.

However, for $p_i\rightarrow + \infty$, while the data stays finite, either the more and more unlikely case of a low signal realization for an increasing variance must have happend, or the more and more unlikely case of a noise canceling the large amplitude signal must have happend.
 
Thus, the a priori as probable case $\tau_i \rightarrow +\infty$ is heavily penalized by the likelihood with respect to the case $\tau_i \rightarrow -\infty$.
Since the PURE filter aims to estimate the mean signal averaged over all $\tau_i $, this imbalance of the likelihood factor lets the regime $\tau_i \rightarrow -\infty$ dominate this average leading to $\langle s \rangle_{(s|d)} =0$.

\subsection{Projection onto generic filter formula}\label{sec:PUREprojected}

We can artificially remove the trivial solution of the PURE filter in case of Jeffreys prior by imposing $dj/dt = 0$ instead of Eq.\ \ref{eq:djdt}. 
This should be understood as looking for a stationary point of the $p$-evolution alone. 
Thus, we are asking for the unique spectrum, which taken as a sharp prior would remain unchanged if a small amount of spectral uncertainty is added. 
This fix point is given by $\beta_i=0$ and therefore
\begin{equation}\label{eq:renormalisedspec}
 p_i = \frac{1+\frac{2}{\varrho_i}}{\varrho_i+1} \mathrm{Tr}\left[ B_i \right].
\end{equation}

Although we have derived this filter only for Jeffreys prior, it is quite plausible to assume that the general spectrum formula, Eq.\ \ref{eq:characteristic equation}, with $(\delta_i,\varepsilon_i)=(1,-0.5/(1+2/\varrho_i))$ also holds for $(\alpha_i, q_i)\neq (1,0)$. We leave a formal proof of this for future work. 
In this form the PURE filter for a Jeffreys prior is projected into the $\delta \epsilon$-plane of the  representation Eq. \ref{eq:characteristic equation} for the MAP filters, which is displayed in Fig.\ \ref{fig:diagram}.

\section{Perception threshold}\label{sec:perception threshold}
\subsection{Critical perception}
\begin{figure*}
 \includegraphics[bb=180 160 650 480,width=\columnwidth]{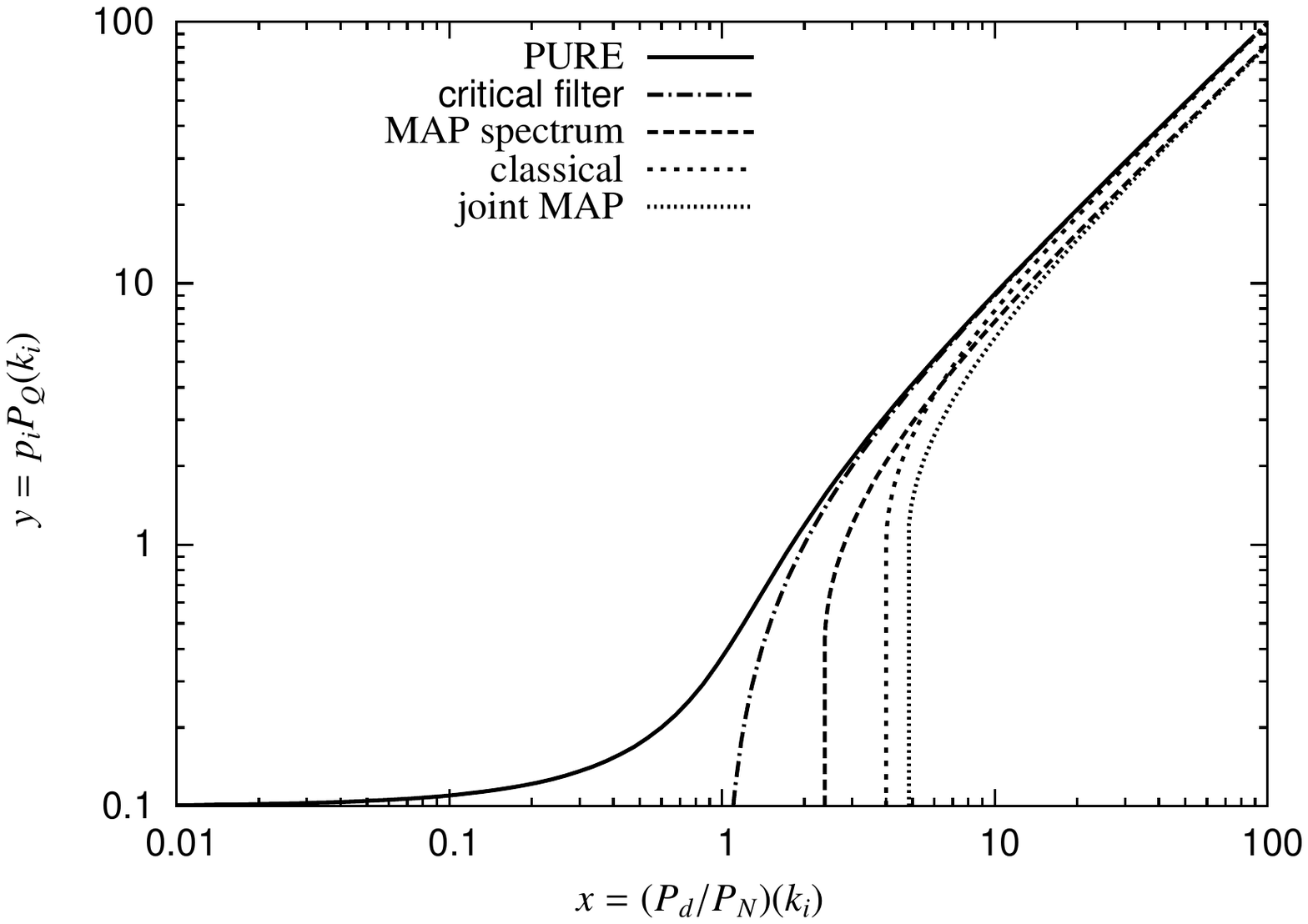}
 \includegraphics[bb=180 160 650 480,width=\columnwidth]{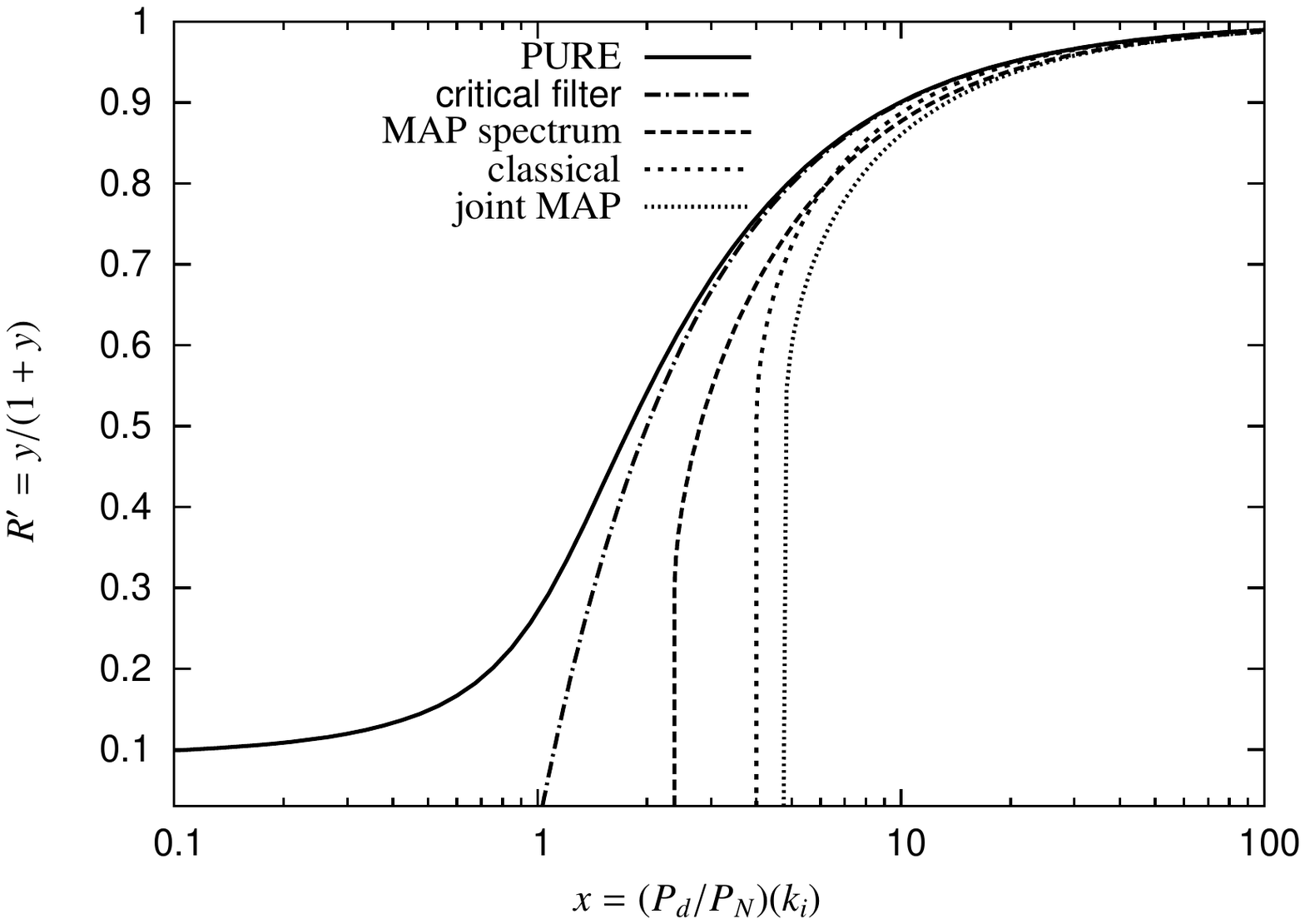}
 \caption{Left: Used spectral power $y$ of the filters as a function of $x$, the data power in noise power units. A spectral bandwidth of $\varrho_i = 8$ was assumed.  Right: The same for the signal pass-through $R'_i$.}
 \label{fig:perceptionTh}
\end{figure*}

In case of Jeffreys prior ($q_i =0$, $\alpha_i=1$, and $t = \infty$), the spectral coefficients $p_i$ used by some of our filters are only non-zero for spectral bands with a data variance above some threshold. 
Bands with lower band power are fully suppressed in the reconstructed map, since the Wiener filter removes completely any fluctuations in bands for which the assumed signal covariance is zero.
Thus, a perception threshold appears for filters within a certain critical line in the $\delta\varepsilon$-plane, which we calculate in the following. 

Filter without perception threshold have to exhibit $p_i>0$, even when the data has no power at all. 
 Thus we investigate the extreme case  $d=0$ by inserting $m_p=0$ into Eq.\ \ref{eq:characteristic equation} and find after some algebra
\begin{equation}
 1 + \frac{2\,\varepsilon_i}{\varrho_i} = \delta_i \, \underbrace{\frac{1}{\varrho_i} \mathrm{Tr}\left( (1+Q_p)^{-1}I_i \right)}_{\le 1},
\end{equation}
with $I_i = S^{-1}_i S_i$ the unit matrix restricted to the $i$-th band. Since the marked expression on the rhs is one only for vanishing $p$, we find the critical line to be given by
\begin{equation}\label{eq:critical line}
 \delta_i^\mathrm{crit} = 1 + \frac{2\,\varepsilon_i}{\varrho_i} .
\end{equation}
Filters with $\delta_i> \delta_i^\mathrm{crit}$ do not exhibit a perception threshold, since even for $d=0$ all $p_i >0$. Filters with $\delta_i < \delta_i^\mathrm{crit}$ exhibit a perception threshold. We note that a non-Jeffreys prior with $\alpha_i>1$ but still $q_i=0$ can also be included into this classification scheme, by just adding $\alpha_i -1$ to $\varepsilon_i$. Filters with $q_i>0$ obviously do not exhibit a perception threshold, since even in the limit of vanishing data and vanishing propagator Eq.\ \ref{eq:characteristic equation} has the positive solution $p_i = q_i/(\gamma_i+\eps_i)$.

The point $(\delta_i, \varepsilon_i)=(1,0)$ lies on top of the critical line, as can be seen in Fig.\ \ref{fig:diagram}, and therefore the term \textit{critical filter} seems to be appropriate for it. 

\subsection{Translation invariant data model}

Here, we calculate the perception thresholds of our filters in the case of a translationally invariant data model. 
Although a general criterion for the position of the threshold in data space can easily be worked out, it is more instructive to investigate a simplified case.
We assume the signal and noise to live in the same spatial space, and their covariances to be fully
characterized by power spectra in Fourier space, 
\begin{eqnarray}
S(k,q) &=& (2\pi)^n\, \delta(k-q) \,P_{S}(k),\nonumber\\
N(k,q) &=& (2\pi)^n\, \delta(k-q) \,P_{N}(k),
\end{eqnarray}
with
$P_s(k) = \langle |s(k)|^2 \rangle /V$, and $P_N(k) = \langle |n(k)|^2
\rangle /V$, where $V$ is the observed volume. We define spectral bands with 
band spectra $P_{S_i}(k)$, so that $P_S(k) = \sum_i p_i \, P_{S_i}(k)$.
We assume further that the signal processing can be completely
described by a convolution with an instrumental beam,
\begin{equation}
 d(x) = \int
dy\, R(x-y) \, s(y) + n(x),
\end{equation}
 where the response-convolution kernel has a
Fourier power spectrum  $P_R(k) = |R(k)|^2$ (no factor $1/V$). 

In this case $D$ can be fully described by a power spectrum,
\begin{eqnarray}
D(k,q) &=& (2\pi)^n\, \delta(k-q) \,P_{D}(k),
\end{eqnarray}
with
$P_D(k) = (P_S^{-1}(k) + P_R(k)\, P_N^{-1}(k))^{-1}$ and all spectral bands decouple.

\subsection{Approximative treatment}

The generic filter equations, Eqs.\ \ref{eq:characteristic equation}, now separate into independent equations for the individual $p_i$. Let us look first at the trace-terms in this equation, which now read
\begin{eqnarray}
\mathrm{Tr}\left[ m_p\,m_p^\dagger S_i^{-1}\right] &=& V \, \int _i \frac{dk}{(2\,\pi)^n} \, \frac{P_d(k) \, { p_i}^2\,P_{Q_i}(k)}{P_N(k) \,(1 + p_i\, P_{Q_i}(k))^2},\nonumber\\
\mathrm{Tr}\left[ D_p  S_i^{-1}\right] &=&  V \, \int _i \frac{dk}{(2\,\pi)^n} \, \frac{p_i}{1 + p_i\, P_{Q_i}(k)}.
\end{eqnarray}
 We define the data power $P_d(k) = |d(k)|^2/V$ and the $i$-band fidelity power $P_{Q_i}(k) = (P_{S_i} \, P_R/P_N)(k)$. We further use the approximation 
$ V\, \int _i {dk}/{(2\,\pi)^n} f(k) \approx \varrho_i \, f(k_i)$, which assumes that $f(k)$, a combination of spectra, does not vary significantly over the narrow spectral band $i$. This permits us to write the generic filter formula, Eq.\ \ref{eq:characteristic equation}, which determines the filter band coefficients $p_i$ as an algebraic and dimensionless expression:
\begin{equation}\label{eq:characteristic}
x = \frac{1+y}{y^2}\, \left[(t\,y -u)\,(1+y) - \delta\, y \right].
\end{equation}
Here, we have dropped the index $i$  and defined the noise-normalized data power $x = P_d(k_i)/P_N(k_i)$ and the measurement fidelity $y = p_i \, P_{Q_i}(k_i)$.
The numerical coefficients are 
\begin{eqnarray}
t &=& \frac{2}{\varrho_i}\,(\gamma_i + \varepsilon_i) = 1 + \frac{2}{\varrho_i}\,(\alpha_i -1 + \varepsilon_i),\nonumber\\
u &=& \frac{2}{\varrho_i}\,q_i\,P_{Q_i}(k_i),\;\mbox{and}\; \delta = \delta_i.
\end{eqnarray}
In case of Jeffreys prior, these simplify to $u=0$ and $t=1 + 2\,\varepsilon_i/\varrho_i$ and the recast generic filter formula Eq.\ \ref{eq:characteristic} has the following solutions
\begin{eqnarray}
 y &=& 0,\;\mbox{and}\\
 y &=&  \frac{x-x_0}{2\,t} \pm \sqrt{\left( \frac{x-x_0}{2\,t}\right)^2  - 1 + \frac{\delta}{t}},
\;\mbox{with}
\nonumber\\
x_0 &=& 2\,t-\delta = 2+4\,\varepsilon_i/\varrho_i -\delta_i.\nonumber
\end{eqnarray}
Although there might be up to three simultaneous real solutions for a given $x$, always the largest value should be taken.
This is in line with our decision to ignore the trivial solution and the expectation that the assumed spectral amplitude $y$ should increase with increasing data power $x$, an not decrease as the lower branch of the square root does.
The largest solution is non-zero only if
\begin{eqnarray}
 x &\ge& x_\mathrm{th} = 
\left\{
\begin{array}{ll}
 0 & x_0 < 1, \\ 
 x_0 + 2 \, \sqrt{t\,(t-\delta)} & x_0 \ge 1.
         \end{array}
\right.
\end{eqnarray}
The assumed dimensionless signal power $y$ is shown in Fig.\ \ref{fig:perceptionTh} as a function of the dimensionless data variance $x$.
Asymptotically, for $x\gg x_0$, we have a linear increase of assumed signal strength and data variance $y(x) = x-x_0$. The critical filter is special in that this relation holds exactly for the full region $x\ge x_\mathrm{th} = x_0$. All of the MAP estimators in this work have $x_\mathrm{th}>x_0$ and exhibit a jump from $y=0$ to $y=\sqrt{1-\delta/t}$
at $x=x_\mathrm{th}$, followed by an approach to the linear asymptotic. The threshold approaches $x_\mathrm{th} \rightarrow 1$ from above for
$\varrho_i \rightarrow \infty $ for the MAP spectrum filter, however, it is always $x_\mathrm{th}=4$ for the classical filter, independent of the spectral bin size $\varrho_i$.

The PURE filter as given in Sec. \ref{sec:PUREprojected} is the only one of our sample, which has no perception threshold  since  $y(x)$ is positive for all $x$.
Even in case the data exhibits negligible variance $x\ll 1$, the filter still uses a non-negligible spectral amplitude, since  $y(0) \approx 1/(\varrho_i+1)$. This might surprise, since the implied assumption of a significant signal variance is obviously not supported by the data. However, the renormalized filter aims for an optimal reconstruction, and not for an accurate power spectrum measurement, and letting some fraction of some data band with apparently low noise realization pass (remember $x\ll 1$) does not spoil this. 

The combination of signal measurement and filtering can be regarded as a single response operator $R'$, with 
$R'\, s= \langle m \rangle_{(d|s)} = F_{p}\, R\, s = D\, M \, s$, which decomposes into separate pass-through factors for the individual bands, 
$R_i' = P_D(k_i) \, P_R(k_i)\,P_N^{-1}(k_i) = y/(1+y)$. 
This is also shown in Fig.\ \ref{fig:perceptionTh}.

\subsection{Consequences for cosmological practice}

The critical filter estimates the power spectrum of a Wiener map, which is (iteratively) filtered with this very same spectrum (until convergence),
while correcting the spectra for an estimate of the filtered-out power during each iteration.
Similar procedures are widely used in cosmology under the names Karhunen-Lo\`eve (KL, \citep{1947KarhunenK,1978LoeveM, 2002MNRAS.335..887T}) and  Feldman-Kaiser-Peacock (FKP,   \citep{1994ApJ...426...23F}) estimators to measure power spectra of galaxy cataloges. 
As the critical filter, these should therefore also exhibit a perception threshold for spectral modes with a data variance not significantly exceeding the noise variance.
Therefore, one would expect that cosmological spectra obtained by these estimator should exhibit modes with zero power. However, in applications of these scheme in the cosmological literature, the iterations of filtering and spectral measurements are usually not repeated until convergence.\footnote{Some random examples: \citet{2002MNRAS.335..887T}, \citet{2007ApJ...657..645P} as well as \citet{1994ApJ...426...23F} use a fixed and constant spectrum in the optimal data weighting step of the KL and FKP schemes, and do not iterate at all.}
Thus the knowledge system keeps some memory of the initial power spectrum choice, which can be regarded as a hidden prior regularizing the spectrum and preventing the perception threshold that a correctly implemented KL or FKP estimator would exhibit (see also \citep{2002MNRAS.335..887T} for a discussion on this).

\begin{figure*}[t]
\includegraphics[bb=125 20 470 375,width=\textwidth]{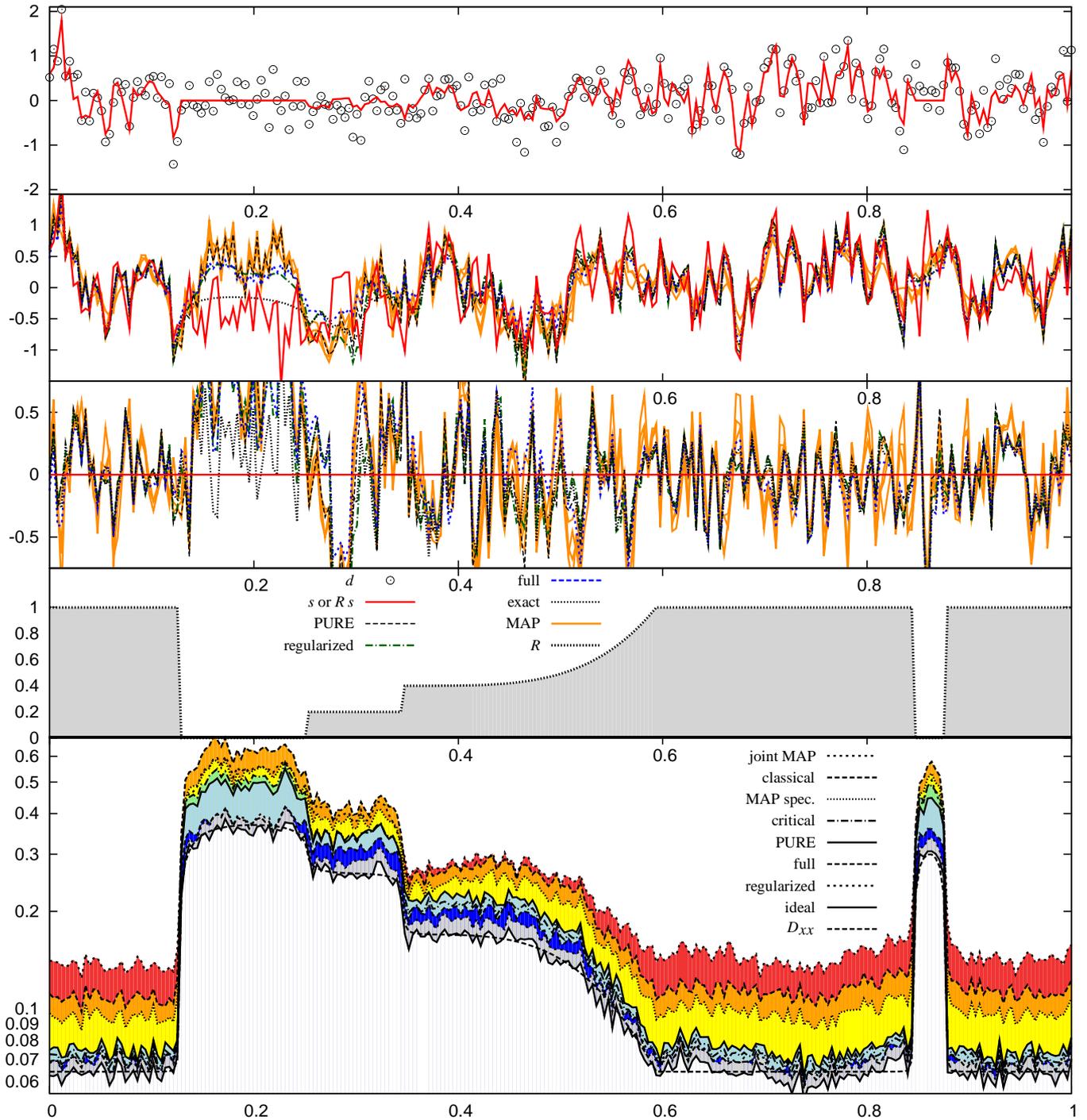}
\caption{First panel: Test data points according to $d = R\,s+n$ and signal response
$R\,s$ in a settign with periodic boundary conditions. Second panel: Signal realization $s$ and the reconstructions as labled in the fourth panel. 
The four MAP reconstructions (joined MAP, MAP spectrum, classical, and critical filter) are shown with the same line 
since they are very similar.
Also the reconstruction using the exact spectrum is displayed. 
Third panel: The same as above, just enlarged and with the signal subtracted to highlight the difference in the  reconstruction errors.
Fourth panel: Response $R$ and line key for the panels above. 
Fifth panel: Error variance $\langle (s_x - m_x)^2 \rangle_{(d,s)}$ of the filters from 700 signal and data realizations in logarithmic units to show the average fidelity of the individual filters.
The order of the line keys reflects roughly the order of the average error of the different methods. The color/grey-scale areas (in online/printed version) should only help to guide the eye.}
 \label{fig:data1}
\end{figure*}
\begin{figure*}[t]
\includegraphics[bb=125 265 473 620,width=1.0\textwidth]{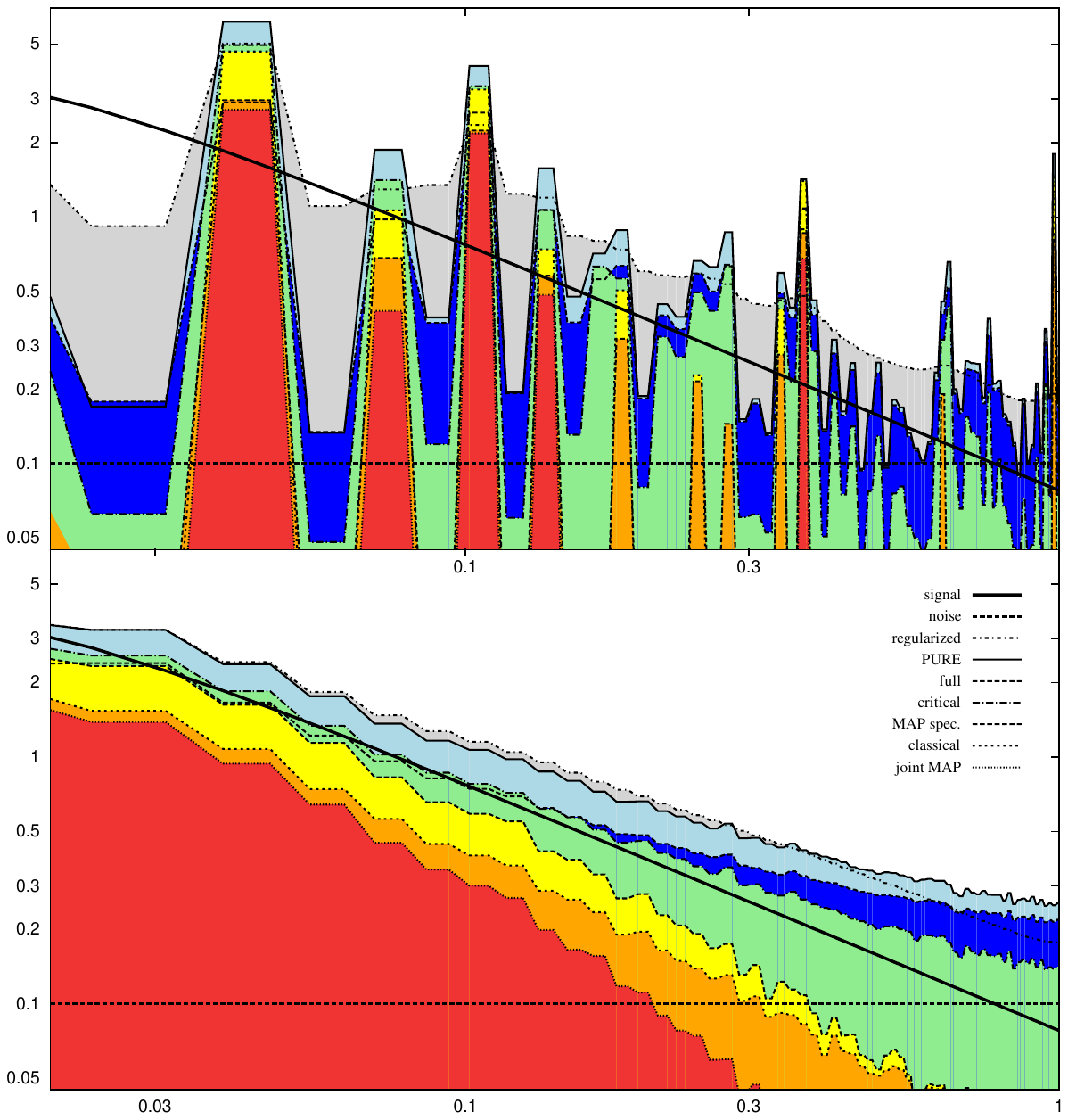}
\caption{Signal and noise spectra in comparison to the assumed spectra of the five filter for the datasets displayed in Fig.\ \ref{fig:data1} in double logarithmic units. $k$-vectors in units of the Nyquist wavevector of $k_{\rm Ny} \approx 256\, \pi$. 
The filter-spectra for the individual dataset of Fig.\ \ref{fig:data1} are shown in the  top panel, the average filter spectra for the 700 signal and data realizations also shown Fig.\ \ref{fig:data1} are displayed in the bottom panel.
The presence of perception thresholds in many of the presented filters is clearly visisible by the many missing frequencies in the top panel and also as the general down-trend of the average spectra close to the crossing of signal and noise spectra.
The order of the line keys reflects the order of the average spectral amplitudes of the different filter at $k=0.3$.}
 \label{fig:spec1}
\end{figure*}

\section{Comparison of the map making algorithms}\label{sec:compare}
\subsection{The test case}

We want to examine the filter performances with an instructive test case. In case the spectral uncertainty is small, all filters in this work can be expected to provide comparable results since they Wiener filter the data with basically the prior spectrum with small differences. Thus, in order to see the differences in performance more clearly, we again adopt Jeffreys prior for our spectral parameters ($\alpha_i=1$ and $q_i=0$, well, for numerical reasons $q_i = 0.01$). A spectrum, which naturally implements this distribution is the famous $1/f$-spectrum, which has equal power per decade in frequency space. To have a finite zero mode and signal variance, we adopt
\begin{equation}
 P_S(k)= P_0 \, (1+(k/k_0)^2)^{-\frac{1}{2}},
\end{equation} 
with $P_0= 5$ and $k_0= 2$. We further assume some white noise with $P_N(k)=\sigma_n^2 = 0.1$. 

In case the response would be constant or a convolution, the spectral inference problems would be separable in Fourier space, as we have shown in the last section. 
In order to have a more complex problem, with coupling between the different unknown spectral parameter, we introduce a non-homogeneous observational signal response $R$ over the 257 pixel of our signal space, as displayed in Fig.\ \ref{fig:data1} together with a test data set. We split the Fourier space in 64 disjunct spectral bands, with $\rho_i=4$ for all but the lowest band, which has $\rho_0=5$, since it also contains the zero mode. Since we are dealing with a real-number signal in a discrete space, we have to take care of the negative frequency spectrum being identical to the positive ones, and therefore our bands are split into identical positive and negative parts, except the zero-band, which is continuous.

The signal reconstructions of the five filters are also shown in Fig.\ \ref{fig:data1}, and the used spectra in Fig.\ \ref{fig:spec1}. The spectra are roughly ordered the way we expect them to be following our perception threshold analysis in Sec.\ \ref{sec:perception threshold}. However, there is the suprising modification that even the renormalized filter seems to suffer from a slight perception threshold, since many of the higher $k$-vector bands with lower signal to noise ratio are nearly free of power. 
A more informative prior for the power distribution would cure this, but this would limit the generality of our filter. So we should look for other yet unexploited prior information.

\subsection{Spectral smoothness regularization}

The $1/f$ signal spectrum adopted in our example is a member of the large class of smooth spectra, which do not exhibit spectral lines, jumps and edges. Spectral smoothness information can easily be incorporated into the framework. Since we do not want to specify a specific smoothness length scale, we require the double logarithmic derivative of the spectrum to be of limited variance. This can be done by introducing an additional prior energy for non-smoothness 
\begin{eqnarray}
 E_\mathrm{reg} &=& 
\frac{1}{2\, \sigma_P^2} \, \int d\log k \,\left( \frac{d\log P_S(k)}{d\log k}\right)^2\\
&\approx& 
\frac{1}{4\, \sigma_P^2} \, \sum_i \frac{k_i+k_{i-1}}{k_{i}-k_{i-1}} 
(\tau_i - \tau_{i-1})^2 \equiv  \frac{1}{2} \,\tau^\dagger T\, \tau. \nonumber
\end{eqnarray}
Here we have (re-)introduced the logarithm of the power spectrum parameters $\tau_i = \log p_i$, have discretized the integral and derivatives, and collected all coefficients in a matrix $T$. The quadratic form in $\tau$ in the last line shows that this is actually a log-normal prior contribution, which can be combined with the log-normal prior appearing in the renormalization calculation. 
Instead of repeating that calculation with now interdependent parameters, we just use our physical intuition to obtain the regularized filter equation for the filter spectrum, and leave any proof or improvement for future work.

The unregularized evolution equation for $\tau$,  Eq.\ \ref{eq:dpidt}, can just be equipped with a regularizing force $-d E_\mathrm{reg}/d\tau$:
\begin{equation}
 \frac{d\tau}{dt} = \beta(\tau) - T\,\tau.
\end{equation}
The regularized Jeffreys prior case is then given by the fix point specified by $\beta(\tau) = T\,\tau$  and reached asymptotically for $t \rightarrow\infty$. The matrix $T$ couples the neighboring bands together and thereby produces much smoother filter spectra without the gaps the other filter spectra exhibit, as can be seen in Fig.\ \ref{fig:spec1}, where the regularized filter spectrum for $\sigma_P = 2$ is shown.

\subsection{Full PURE filter}

Spectral smoothness can not always be assumed, and therefore we should also think of other ways to improve the filter fidelity. One way is to be more precise in the  PURE filter derivation. The largest approximation made was probably the neglection of the $dj/dt$ term, which for infinite spectral uncertainty, $t\rightarrow \infty$, leads to a trivial solution of $m=0$. If we want to include this term, we can therefore only apply it for a finite amount of uncertainty, say up to $t=1$. This implies that the initial starting point of the spectral renormalization flow would influence our result. In case of a concrete application, this might be very desirable, since there a good initial guess for the spectrum might be available. 

In our more abstract discussion here, we want to avoid such choices, also in order to be sure not to have included too much spectral prior knowledge into the filter preventing a fair comparison to the others. Therefore we start the renormalisation flow including the $dj/dt$ term with the fix point spectrum of the approximated PURE filter (without this term) and stop it at $t=1$. This way we have both, independence of any prior spectrum and inclusion of non-Wiener corrections. The resulting filter seems to be partly cured from too generous predictions in regions without data while the results in better determined regions are practically unchanged, as can be seen in Fig.\ \ref{fig:data1}.

This can be understood in the following way. We have roughly $dj/dt \propto -S^{-1}m$, since $m^\dagger S_i^{-1}m \approx  \varrho_i$ for most modes. If there is power at a poorly observed location in the map  $m$ on a level comparable to the well observed ones, $j$ evolves in both regions with similar speed. However, the effect of this evolution to the map $m = D\,j$ is larger in regions with larger uncertainties, since $D$ is larger there. Thus, any power spilled into observational gaps is removed faster than power in well observed regions. The full PURE filter seems to be aware of the lower certainty of the former.

\subsection{Statistical comparison}

A statistical assessment of the different filters is also shown in Fig.\ \ref{fig:data1}. There it is apparent that the filters derived from MAP principles are worse than the PURE filter, with only the critical filter being comparable in performance. The underestimation of the power spectra due to the perception threshold obviously reduces the fidelity of those filters. 

The spectral smoothness regularized, renormalized filter clearly outperforms the unregularized ones, probably due to the lack of spectral gaps. Its performance is comparable to that of the Wiener filter using the correct signal power spectrum $P_S(k)$. 
The error variance for the latter filter is also displayed in Fig.\ \ref{fig:data1} in comparison to its theoretical value given by the Wiener variance $D_{xx}$ (see Eq.\ \ref{eq:mapuncertainty}). 
Finally, also the full PURE filter as described in the last section is shown. 
Its fidelity is comparable to the spectrally regularized one, without that any spectral smoothness assumptions had to be made. 
Of course, such assumptions could also be included into this filter.

\section{Conclusions}\label{sec:conclusion}
We showed how to deal with parameter uncertainties in information field theory by introducing an effective Hamiltonian over the joint space of the signal field and the parameters. In order to go beyond a classical, or Maximum a Posteriori treatment of the problem we presented an uncertainty renormalization scheme, in which the parameter uncertainty is successively fed into the knowledge system. The resulting parameter uncertainty renormalized estimation, PURE, can be used to tackle many signal inference problems including calibration uncertainties.

It seems that the PURE provides a Gaussian approximation to the full posterior probability function, which has maximal cross-information with it, as thermodynamic considerations in \cite{2010PhRvE..82e1112E} have shown.

To demonstrate the advantage of PURE with a concrete example, we investigated the general problem of inferring a Gaussian signal with unknown spectrum from noisy data, which follows from a linear, but inhomogeneous data model. Following the parameter uncertainty renormalization and various classical approaches, four classical and one renormalized filter were derived. All filters can be regarded as Wiener filter operations with assumed signal spectra to be calculated from the data by a single recipe, Eq.\ \ref{eq:characteristic equation}, with just differences in two of its numerical coefficients.

The computational complexity of all those filters is therefore very similar and should not be a reason to prefer one over the other. Their signal fidelity, however, differs significantly. 
In case a non-informative Jeffreys prior is adopted for the spectral amplitudes, all classical filters suffer from a perception threshold.
Spectral bands, which do not show more data power than the threshold, are completely filtered out. Three out of the four classical filters investigated have a perception threshold which requires data variance significantly above the noise level. The fourth one, the critical filter, lives on the critical line between filters with and without perception threshold in our space of filter parameters.  The critical filter tries to match the correct spectrum on a logarithmic power scale. Its perception starts therefore for modes with a variance just above the noise level, as soon the data indicates some potential signal power.
It has recently been applied successfully to the reconstruction of an all sky map of the galactic Faraday depth \citep{2010arXiv1008.1246O}.

The critical filter coresponds in general to the Karhunen-Lo\`eve method \cite{1947KarhunenK,1978LoeveM, 2002MNRAS.335..887T}, and for an infinite window function to the FKP method \cite{1994ApJ...426...23F} frequently used in cosmology to estimate power spectra of galaxy catalogs. It seems that the perception threshold of this method is often `cured' in applications by a truncation of the full iterative scheme. This implies the presence of a hidden spectrum prior in such estimates.

The PURE filter precepts also for spectral bands, which by chance exhibit less power than expected for the noise alone.
This might appear as being too generous -- the signal spectrum adopted by this filter is typically larger than the correct and therefore optimal, but unknown one. 
However, the PURE filter exhibits the largest fidelity of our filter sample, even slightly better than that of the critical filter. 
The reason lies in the asymmetric fidelity loss for under- and overestimating the true signal spectrum. 
Spectrum underestimation is much worse than overestimation in terms of signal reconstruction accuracy. 
The renormalized filter knows about this and adds a safety margin to any spectral band. 
This margin is inversely proportional to the number of data degrees of freedom informing about the signal spectrum in this band. 
Thus, in the limit of a large number of data points determining the band spectrum the renormalized filter approaches the critical one, but always from the perception threshold free side.

Although the classical filter resulted from maximizing the exact effective, parameter marginalized Hamiltonian (Eq. \ref{eq:MAPmapHamilton}), it performs much worse than the critical and PURE filters. 
Thus, this is an example where the MAP principle, or equivalently a tree-level IFT calculation, provides a poorly performing algorithm, and uncertainty loop corrections as explicitely included in the PURE filter or even the critical filter are essential. 

The PURE filter, as well as the others, can be further improved by adding any additional spectral information. One way is to use informative priors on the spectral behavior, which instantaneously cure the perception threshold problem. However, even in case no information on the location of the spectrum is available, information about its smoothness as a function of the Fourier space coordinate may be exploited. We show that the performance of the PURE filter with spectral smoothness prior approaches that of the optimal Wiener filter for known signal power spectrum.

Since the computational complexity of the renormalized filter is identical to the critical one already used in cosmology, there exists no reason not to use it for Wiener filtering of signals with unknown spectra. One only has to keep in mind that the internally used spectrum of the filter is not the best estimate of the signal spectrum, but an overestimate. The critical spectrum provides such an estimate, using the posterior maximum for the logarithm of the spectral amplitudes.

The full PURE filter, which contains non-Wiener filter corrections and requires the more expensive evaluation of the renormalization flow equation, performs best among all spectrally unregularized filters. Spectral smoothness information can also be incorporated into it if available. 

To conclude, the PURE scheme to construct optimized filters presented in this work is very general and should also be applicable to the problems of inference with uncertainties in the instrument response, the typical calibration problem, and for measurements without known noise level. A better understanding of the implications and assumptions of the commonly used process of self-calibration should be feasible, and possibly also improvements thereof. The pseudo-time parameter appearing in the renormalization flow, the amount of uncertainty or parameter dispersion fed into the knowledge system, may be connected under certain circumstances to real physical time. For measurement devices with drifting calibration or noise parameters, and also for signals with a slow, but unknown  time evolution of their signal spectra, the parameter uncertainty renormalization equation offers a natural possibility to model this. Once the amount of uncertainty dispersion per physical time is fixed, the equation permits to continuously update the unknown parameters by combining past and novel information in an optimal, and controlled way. The PURE approach may thereby make contributions to the technologically important field of optimal control and time dependent instrument calibration.

\section*{Acknowledgments}
We acknowledge helpful discussions with and comments on the manuscript by Jens Jasche, Henrik Junklewitz, Cornelius Weig, Marco Selig, Niels Oppermann, Gerhard B{\"o}rner, Benjamin Wandelt, and three anonymous referees.

\appendix

\section{Signal covariance likelihood}\label{sec:signal covariance like}

In order to find the posterior of the signal covariance, we have to
calculate $P(p)\,Z_p/Z$. 
We show below that the evidence $Z_p[0]$ for any
parameter $p$ of the free Hamiltonian, given by Eq.\ \ref{eq:ZdfreeTheory}, is
\begin{equation}\label{eq:dataLikelihooparam}
 Z_p = P(d|p) = \G(d,R\,S_p\,R^\dagger + N ).
\end{equation}
This formula can also intuitively be read as the data likelihood given $p$,
since it compares the power in the data to their expected fluctuations level
$\langle d\,d^\dagger \rangle_{(d,s|p)} = R\,S_p\,R^\dagger + N$. 
It can therefore be used for a Bayesian estimate of any model-parameter of the free theory, not only for spectral parameters as in this work.

Proofs for  Eq. \ref{eq:dataLikelihooparam} can be found in  \citep{2008MNRAS.389..497K, 2008MNRAS.391.1315F}. 
However, these proofs rely on either on the very special assumption fo $R$ being invertible \citep{2008MNRAS.389..497K} or on a Taylor expansion of the logarithm of a marix \citep{2008MNRAS.391.1315F}, which has actually a limited convergence radius and therefore is not sufficient for a general proof. 
A proof without such limitations goes as follows: 
First, we concentrate only on the dependence of $P(d|p)$ on the data $d$,
\begin{eqnarray}
Z_p &=& P(d|p) = \int \!\mathcal{D}s\, P(d,s|p)\nonumber\\
&\propto &  \!\!\! \int \!\mathcal{D}s  \exp\!\left(\! -\frac{1}{2}\! \left( s^\dagger S_p^{-1} s + (d-R\,s)^{\dagger} N^{-1} \, (d-R\, s)\right)\! \right) \nonumber\\
&\propto &  \exp \left(- \frac{1}{2} \left(d^\dagger  (N^{-1} -  N^{-1}R\, D_p R^\dagger N^{-1}) \, d \right) \right)
\nonumber\\
& \propto& 
\exp \left(- \frac{1}{2} d^\dagger (R\,S_p\,R^\dagger + N )^{-1} d\right)\nonumber\\
& \propto& \G(d,R\,S_p\,R^\dagger + N ).\nonumber
\end{eqnarray}
Here, we used $D_p = (S_p^{-1} + M)^{-1}$ with $M = R^\dagger N^{-1} R$.
The second last step relied on $R\,S_p\,R^\dagger + N$ being the inverse of $N^{-1} -  N^{-1}R\, D_p R^\dagger N^{-1}$:
\begin{eqnarray}
(R\,S_p\,R^\dagger + N)\, (N^{-1} -  N^{-1}R\, D_p R^\dagger N^{-1})  &=& \nonumber\\
R\,(S_p - S_p\,M\, D_p - D_p)  R^\dagger N^{-1} + 1 &=&\nonumber\\
R\,(S_p - S_p\,(D_p^{-1} -S_p^{-1})\, D_p - D_p)  R^\dagger N^{-1} + 1 \nonumber
&=&1.
\end{eqnarray}
Second, we have to show that $Z_p$ has the same normalization the Gaussian in Eq. \ref{eq:dataLikelihooparam} has.  
This is most easily seen by
\begin{eqnarray}
\int \! \mathcal{D}d\, Z_p &=& \int \!\mathcal{D}d  \int \!\mathcal{D}s\, P(d,s|p)\nonumber\\
&=&\int \!\mathcal{D}s\, P(s|p)  \int \!\mathcal{D}d  \, P(d|s) \nonumber\\
&=& \int \!\mathcal{D}s\, \G(s, S_p)  \int \!\mathcal{D}n \, \G(n, N)= 1,\nonumber
\end{eqnarray}
where in the last line we replaced the data space integration variable $d $ by a linear shift with the noise variable $n = d- R\, s$ and used the fact that Gaussians are normalized to unity. 
Thus,  Eq. \ref{eq:dataLikelihooparam} is proven.

\section{Derivation of the Gaussian prior}
\label{sec:central limit theorem}

Here we show how the different auxiliary variables $\tau_{ij}$ combine into a normal distribution for $\tau_i = \sum_j \tau_{ij}$, as was assumed in Sec.\ \ref{sec:lognormal prior}. We drop in the following the index $i$, which labels the signal bands. Since we assume $e^{\tau_{j}}$ to be distributed according to Eq.\ \ref{eq:p-prior},
we have
\begin{equation}
 P(\tau_j) = \frac{\exp \left[ {-(\alpha -1)\, (\tau_j -\log q) - q\, e^{-\tau_j}} \right] }{\Gamma[\alpha -1]} 
\end{equation}
The non-bias condition, Eq.\ \ref{eq:unbiased auxiliary priors}, translates into
\begin{equation}
 \langle \tau_j \rangle_{(\tau_j)} = \log q - \psi_0(\alpha-1)   = 0,
\end{equation}
with $ \psi_n(z) $ being the Polygamma function. 
This condition fixes $q(\alpha) = e^{ \psi_0(\alpha-1) }$, which for large values of $\alpha$, and thereby for well localized auxiliary parameters, is asymptotically $q = \alpha - \frac{3}{2}$. 
The dispersion of the auxiliary variables is 
\begin{equation}
 \delta t = \langle \tau_j^2 \rangle_{(\tau_j)} =  \psi_1(\alpha-1),
\end{equation}
which asymptotically is $\delta t=1/(\alpha -1)$ for large $\alpha$.

Now, we can work out the total prior resulting from the combination of $N =t/\delta t$ auxiliary variables, where $t$ is the uncertainty level of the prior, and $\delta t$ that of the individual variables:
\begin{eqnarray}
 P(\tau)  \!\!\!\! \!\!&=&  \!\! \!\! \!\!\left( \prod_{j=1}^N \int\! d\tau_j \, P(\tau_j)\right) \, \delta(\tau - \sum_{j=1}^N \tau_j)\nonumber\\
 \!\! \!\! \!\!&=& \!\!   \!\!\!\! \int\! \frac{dk}{2\pi}\,\left( \prod_{j=1}^N \int\! d\tau_j \, P(\tau_j)\right) \, e^{-i\,k\,(\tau- \sum_{j=1}^N \tau_j)}\nonumber\\
 \!\! \!\! \!\!&=&  \!\!\!\!  \!\! \int\! \frac{dk}{2\pi} \left( \int\!  \frac{d\tau_j \,q^{\delta t^{-1}}\,\exp \left[ {-(\delta t^{-1} -i\, k)\, \tau_j  - q\, e^{-\tau_j}} \right] 
}{\Gamma[\delta t^{-1}]} \right)^N \!\!   \!\! \!\!\nonumber\\
&\times & e^{-i\,k\,\tau}\nonumber\\
 \!\!  \!\!\!\!&=& \!\!  \!\! \!\! \int\! \frac{dk}{2\pi}\,\left(  \frac{\Gamma[\delta t^{-1} - i\, k]}{\Gamma[\delta t^{-1}]}\, q^{i\,k}\right)^N e^{-i\,k\,\tau}\nonumber\\
 \!\!  \!\!\!\!&=& \!\!  \!\! \!\!\int\! \frac{dk}{2\pi}\,\exp\left[ {-i\,k\,\tau} + N\,\log \left(\frac{\Gamma[\delta t^{-1} - i\, k]}{\Gamma[\delta t^{-1}]}\right) \right. \nonumber \\
&&\,\,  \,\, \,\,\,\, \,\,\,\, \,\, \,\,\,\, \,\,\,\,\,\,\left.
-i\,k\,N\,\psi_0(\alpha-1) 
\begin{array}{c}
\\ \\
\end{array}\!\!\!\!
\right]  \nonumber\\
 \!\! \!\! \!\!&=& \!\!  \!\! \!\!\int\! \frac{dk}{2\pi}\,\exp\left[ {-i\,k\,\tau} - \frac{t}{2}\,k^2  + \mathcal{O}(\delta t\,k)\,k^2 \right] \, \nonumber\\
 \!\! \!\!&\longrightarrow &  \G(\tau,t)\;\mbox{for}\; \delta t\rightarrow 0,
\end{eqnarray}
as also expected from the central limit theorem of statistics. Thus, the resulting distribution for the $p_i$ parameter is log-normal.

\bibliography{bibtex/ift}
\bibliographystyle{bibtex/aa}

\end{document}